\documentclass[letter,scriptaddress,twocolumn, prl,showkeys,showpacs,superscriptaddress,longbibliography]{revtex4-1}

	\usepackage{amsmath}
	\usepackage{makeidx}
	\usepackage{algorithm,algcompatible}
\algnewcommand\INPUT{\item[\textbf{Input:}]}%
\algnewcommand\OUTPUT{\item[\textbf{Output:}]}%
    \usepackage{booktabs}
	\usepackage{amsfonts}
	\newcommand\numberthis{\addtocounter{equation}{1}\tag{\theequation}}
	\usepackage{graphicx, xcolor}  
	\usepackage{dcolumn}   
	\usepackage{bm}        
	\usepackage{amssymb}   
	\usepackage[ansinew]{inputenc}
	\usepackage[usenames,dvipsnames]{pstricks}
	\usepackage{multirow}
	\usepackage{caption}
	\usepackage{epstopdf}
	\usepackage{pst-grad} 
	\usepackage{pst-plot} 
	\usepackage[colorlinks,hyperindex]{hyperref}
	\hypersetup
	{
		colorlinks,%
		citecolor=black,%
		linkcolor=black,%
		urlcolor=black,%
	}
	\usepackage{bbm} 
	\usepackage[font=footnotesize,labelfont=bf]{caption}
	
\newcommand{\be}{\begin{equation}}
\newcommand{\ee}{\end{equation}}
\newcommand{\ba}{\begin{aligned}}
\newcommand{\ea}{\end{aligned}}
\newcommand{\lp}{\left(}
\newcommand{\rp}{\right)}

\begin{document}

\title{Mechanics of allostery: contrasting the induced fit and population shift scenarios}
\author{Riccardo Ravasio}

\affiliation{Institute of Physics, \'Ecole Polytechnique F\'ed\'erale de Lausanne, CH-1015 Lausanne, Switzerland}

\author{Solange Marie Flatt} 
\affiliation{Institute of Physics, \'Ecole Polytechnique F\'ed\'erale de Lausanne, CH-1015 Lausanne, Switzerland}
\author{Le Yan}
\email{Current address: Chan Zuckerberg Biohub, 499 Illinois St, San Francisco, CA 94158}
\affiliation{Kavli Institute for Theoretical Physics, University of California, Santa Barbara, CA 93106, USA}
\author{Stefano Zamuner}
\affiliation{Institute of Physics, \'Ecole Polytechnique F\'ed\'erale de Lausanne, CH-1015 Lausanne, Switzerland}
\author{Carolina Brito}
\affiliation{Instituto de F\'isica, Universidade Federal do Rio Grande do Sul, CP 15051, 91501-970 Porto Alegre RS, Brazil}
\author{Matthieu Wyart}
\email{Corresponding author: riccardo.ravasio@epfl.ch, matthieu.wyart@epfl.ch}

\affiliation{Institute of Physics, \'Ecole Polytechnique F\'ed\'erale de Lausanne, CH-1015 Lausanne, Switzerland}


\begin{abstract}

In allosteric proteins, binding a ligand can affect function at a distant location, for example by changing the binding affinity of a substrate at the active site. The induced fit and population shift models, which differ by the assumed number of stable configurations, explain such cooperative binding from a thermodynamic viewpoint. Yet, understanding what mechanical principles constrain these models remains a challenge. Here we provide an empirical study on 34 proteins supporting the idea that allosteric conformational change generally occurs along a soft elastic mode presenting extended regions of high shear. We argue, based on a detailed analysis of how the energy profile along such a mode depends on binding, that in the induced fit scenario there is an optimal stiffness $k_a^*\sim 1/N$ for cooperative binding, where $N$ is the number of residues.  We find that the population shift scenario is more robust to mutation affecting stiffness,  as binding becomes more and more cooperative with stiffness up to the same characteristic value $k_a^*$, beyond which cooperativity saturates instead of decaying. We confirm numerically these findings in a non-linear mechanical model. Dynamical considerations suggest that a stiffness of order $k_a^*$ is favorable in that scenario as well, supporting that for proper function proteins must evolve  a functional elastic mode  that is softer as their size increases. In consistency with this view, we find a \textcolor{black}{fair} anticorrelation between the stiffness of the allosteric response  and protein size  in our data set.

\end{abstract}

\maketitle

\textbf{Significance statement: }Many proteins are allosteric: binding a ligand affects their activity at a distant site. Understanding the 
principles allowing for such an action at a distance is both of fundamental and practical importance. From the thermodynamic viewpoint, two models have been proposed, according to which binding creates a new stable configuration or  instead shifts the thermal equilibrium between existing states.  We perform a mechanical analysis of these models and show that they are not equally robust to mutations. We argue that in both cases function can most properly occur along a soft elastic mode, whose stiffness decreases rapidly with protein size. We show data  on 34 proteins substantiating this result, supporting a new principle for allosteric design.
\newline

Many proteins are allosteric: binding a ligand at one or several allosteric sites can regulate function at a distant site,  a long-range communication often accompanied by  large conformational changes \cite{Perutz60,Perutz70}. 
There is a considerable interest in predicting the amino acids involved in this communication, or ``allosteric pathway'', from structure or sequence data \cite{Halabi09,Amor16}, since they can be used as targets for drug design \cite{Nussinov13}. 
Yet, understanding the physical principles underlying such action at a distance in proteins remains a  challenge \cite{Wodak19,Thirumalai18}. 
From a thermodynamic standpoint, two distinct views have been proposed. 
In the induced fit scenario,  exemplified by the Koshland-N\'emethy-Filmer (KNF) model \cite{Koshland66}, the protein essentially lies in one single state. 
The latter changes as binding occurs, leading to a  conformational change. In an energy landscape picture such as that of Fig.\ref{fig:sketch_intro}B, it corresponds to a displacement of the energy minimum upon binding.
By contrast, in the population shift model,  exemplified by the Monod-Wyman-Changeux (MWC) model \cite{Monod65}, two states are always present. 
Their relative stability can change sign upon binding, leading to an average conformational change.
Although each of these models presumably  applies to various proteins, they do not specify which designs  allow for efficient action at a distance, and how robust these designs are to mutations \cite{Hopfield73}.

In several proteins~---~see  below for a systematic study~---~it has been observed that the allosteric response induced by binding  occurs predominantly along one or few vibrational modes  \cite{Kitao99,Bahar99b,Xu03,Rios05,Zheng06}. 
This result supports that in at least some proteins elasticity ~---~possibly non-linear~---~ is an appropriate language to describe allostery (in contrast to intrinsically disordered proteins that may be considered more as liquids than solids, for which the  analysis proposed here would not hold).  
Very recently, there has been a considerable effort to use in-silico evolution \cite{Hemery15,Tlusty16}  to study how linear elastic materials can evolve to accomplish an allosteric task \cite{Yan17,Rocks17,Flechsig17,Yan18,Dutta18,Wang18,Flechsig18,Bravi18,Rocks19}. 
In general, binding a ligand locally distorts  the protein, which is modelled by imposing  local displacements at some site, generating an extended elastic response  that in turn determines fitness (chosen specifically to accomplish a given task). 
These models fall into the induced fit scenario, since in the framework of linear elasticity there is always a single  minimum of energy. 
A particularly key allosteric function within proteins is the amount of cooperative binding, defined as the change of binding energy of a substrate at the active site caused  by binding a ligand at the allosteric site. 
Materials optimized to display such a cooperativity over long-distances develop  a single extended ``mechanism''~---~a soft elastic mode, such as the motion of closing scissors~---~connecting the two binding sites \cite{Yan18}. 
It is found that the stiffness $k_a$ (i.e. the curvature of the energy) of this mode cannot be too large nor too small for cooperativity to occur, and that optimal design corresponds to  $k^*_a\sim 1/N$ \textcolor{black}{where $N$ is the number of particles in the system}. If proteins are nearly optimal, mutations stiffening that mode should thus diminish cooperativity. 
Yet, these predictions are restricted to strictly linear elasticity, an approximation that presumably does not hold in the regime where most protein operate~---~certainly not in the population shift scenario.

In this work, we show that the population shift and the induced fit models are very different from a mechanical perspective. 
In the former case as well, function can be achieved by developing a mechanism or soft elastic mode, but cooperativity, instead of steadily decreasing, saturates to a constant value once the  mode stiffness passes some characteristic value, \textcolor{black}{whose scaling with $N$ is $k^*_a\sim 1/N$}. We confirm this prediction in a non-linear elastic model of allostery. This result implies  that cooperativity is  more robust towards mutations increasing stiffness in the population shift scenario. Yet, displaying a stiffness much larger than $k^*_a$ implies a very long transition time between the two states, and is presumably prohibited, suggesting the hypothesis that allosteric proteins function with modes presenting a stiffness near the characteristic value $k_a^*$ in both cases. We test this proposition systematically using X-ray crystallographic data of  34 allosteric proteins. We first confirm that one or a few vibrational modes contribute to the allosteric response, and introduce a new observable establishing  that this response presents unusually extended regions of large shear, as found previously  for three proteins \cite{Mitchell16}. Next we confirm that  the characteristic mode stiffness tends to decrease with the propagation length as we expect \textcolor{black}{from the predicted scaling of $k_a^*$}. Finally, we suggest systematic mutational studies to further test how mechanics constrains allostery.

\section*{Geometrical interpretation of mechanical constraints in induced fit allostery} 

\begin{figure}[h]
 	\centering\includegraphics[width=8.2cm]{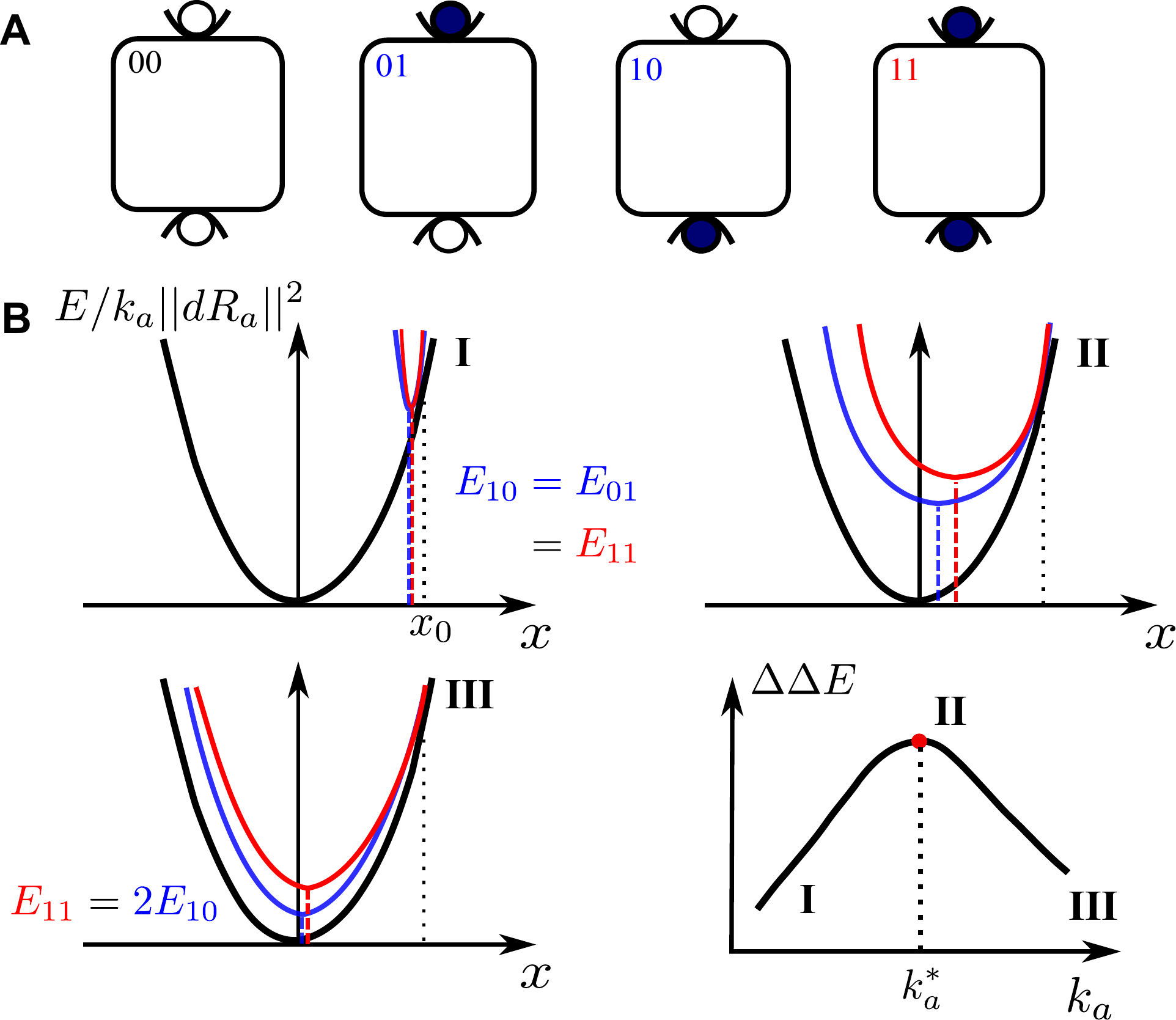}
 	\caption{(A) A protein with two binding sites can be unbound ``00", bound to a single ligand ``01" (or ``10") or doubly bound ``11".
 	 (B) Induced fit scenario. Elastic  energy of the unbound state $E_{00}(x)/k_a$ (in black) re-scaled for visibility by the soft mode stiffness $k_a$, singly bound state $E_{01}(x)/k_a$ (in blue) and doubly bound state $E_{11}(x)/k_a$ (in red) as a function of the imposed motion $x$ along its soft mode.    $E_{00},E_{10}$ and $E_{11}$ correspond to the minima of the black, blue and red curves respectively. In this sketch we have assumed that for a motion $x_0$, the protein shape can accommodate perfectly both ligands without deforming, thus the three energy profiles are identical at that point. Three cases are sketched, depending on the magnitude of the characteristic stiffness of the mode $k_a$.
 	 }
 	 \label{fig:sketch_intro}
 \end{figure}

As sketched in Fig.\ref{fig:sketch_intro}A, a protein with two binding sites can be unbound (labeled ``00"), bound to a single ligand (labeled ``01" or ``10") or doubly bound (labeled ``11"). We define by $E_{00},\, E_{10},\,E_{01}, \,E_{11}$ the  energy of the protein in these four situations (corresponding to the minimum energy of their energy landscape), and choose  $E_{00}=0$ as the reference  energy. Cooperativity is then defined as:
\begin{equation}
    \Delta\Delta E=E_{10}+E_{01} - E_{11}
\end{equation} 
To simplify notation below, we assume a symmetry between the states 01 and 10, in particular $E_{10}=E_{01}$. Our qualitative conclusions below however remain valid even if this symmetry does not hold. 

Consider a  protein displaying cooperativity thanks to the presence of a soft elastic mode. 
Let us denote by $x$ a variable indicating the motion along that mode (see our numerical model below for a concrete example in the context of a shear design),
which varies from zero to unity as the protein undergoes its allosteric response. 
The energy profile $E_{00}(x)$ of the unbound state follows:
\begin{equation}
E_{00}(x)= k_a ||d R_a||^2 f(x)
\end{equation}
where $||d R_a||^2=\sum_{i=1...N} ||d{\vec  R_i}||^2$ is the norm of the allosteric response $|dR_a\rangle \equiv \{ d{\vec  R_a(i)}\}$ and $d{\vec  R_a(i)}$ is the vector displacement of the amino acid $i$. We expect that if the protein has $N$ residues, $||d R_a||^2$ is of order $N a^2$,  where $a$ is the inter-atomic distance. We will confirm empirically the dependence of $||d R_a||^2$ with $N$ below.  $k_a$ is the mode stiffness and $f(x)$ some function of order unity such that $f''(x=0)=1$. For a purely linear elastic material, $f(x)=x^2/2$. More generally for the induced fit scenario, $f(x)$ presents a single minimum, as illustrated in Fig.\ref{fig:sketch_intro}B. In \cite{Yan18} it was argued that if linear elasticity applies, then there is an optimal stiffness $k_a^*$ for cooperativity.
We present a geometrical interpretation of this result, which extends it to the induced fit scenario in general, and will be useful to explain why the population shift scenario behaves differently. Local chemical interactions which lead to identical binding energies for the  inactive and active states do not affect cooperativity, and thus need not be considered here. For concreteness, we assume that after some motion $x_0=1$ along that mode, the protein shape can accommodate perfectly both ligands without deforming. It implies that the  energy profile of the bound states 
$E_{10}(x)$ and $E_{11}(x)$ satisfy $E_{10}(x_0)=E_{11}(x_0)=E_{00}(x_0)$, as pictured in Fig.\ref{fig:sketch_intro}B.
However as $x$ departs from $x_0$, the protein shape does not match the ligands, imposing an elastic strain at the binding sites leading to an increase of elastic energy in the protein, that will trigger motion along other elastic modes. Assuming that the ligands are rigid and that each binding site involves on the order $n_0$ atoms that  move  by a distance $a$ as the protein undergoes its allosteric response, \textcolor{black}{we have $E_{01}(x)-E_{00}(x)= n_0 k a^2 g(x-x_0)$ where $k$ characterizes the stiffness of inter-atomic interactions and is large in comparison with the one of soft  modes, i.e. $k\gg k_a$. Here $g$ is some dimensionless function vanishing quadratically in zero~---~but possibly non-linear at large arguments.}   

$E_{11}$, $E_{01}$ can be computed as the minimum of the curves $E_{11}(x)$ and $E_{01}(x)$ respectively, from which the cooperativity $\Delta \Delta E$ is readily computed. Two extreme  cases occur, illustrated in Fig.\ref{fig:sketch_intro}B:

(i) if $k_a N\ll n_0 k$  (\textcolor{black}{Fig.\ref{fig:sketch_intro}B.I}), as $x$ moves away from $x_0$, the elastic energy induced by binding $n_0 k a^2g(x-x_0)$ is very significant in comparison to the mode energy $E_{00}(x)\sim k_a N a^2 f(x)$. Thus both $E_{10}(x)$ and $E_{11}(x)$ have a sharp minimum near $x_0$, with $E_{11}\approx E_{10}\approx k_a ||d R_a||^2 f(1) $. Thus  $\Delta\Delta E=E_{10}+E_{01} - E_{11}\approx E_{10}\approx k_a ||d R_a||^2 f(1)$, which vanishes as $k_a\rightarrow 0$. 

(ii) If $k_a N\gg n_0 k$ is very large (\textcolor{black}{Fig.\ref{fig:sketch_intro}B.III}), $n_0 k a^2g(x-x_0)$ is small in comparison to $E_{00}(x)$: $E_{00}(x)$, $E_{11}(x)$ and $E_{01}(x)$ are very close to each other, and must thus all present a minimum near $x=0$. Thus binding does not trigger motion along the soft mode, whose presence is useless. No extended modes couples the two binding sites and $E_{11}\approx E_{01}+E_{10}$, leading to $\Delta\Delta E \rightarrow 0$ as $k_a\rightarrow \infty$. 

(iii) Optimal cooperativity is thus found at some intermediary $k_a^*\sim n_0 k/N$, corresponding to \textcolor{black}{Fig.\ref{fig:sketch_intro}B.II}. Note that the present argument for an optimal $k_a^*$ does not require the energy profile $f(x)$ to be an exact parabola, as long as it is monotonically growing in both directions around its minimum.  

\subsection*{Mechanical aspects of the population shift model}

\subsubsection*{MWC model}
We recall some aspects of the MWC model. To simplify notation, we consider that the protein displays two symmetric binding sites  as illustrated in Fig.\ref{fig:sketch_intro}A. 
The protein is assumed to lie in two possible distinct configurations, ``Inactive'' ($In$) and ``Active'' ($Ac$). In the absence of binding, we take the  energy  of the inactive state as our reference  (i.e. $E_{00}^{In}=0$) and  denote the energy of the active state  $E_{00}^{Ac}=E_0$.  We assume that the active configuration has a well-suited geometry to bind each ligand, thus no elastic energy is spent for binding and $E_{10}^{Ac}=E_{01}^{Ac}=E_{11}^{Ac}=E_0$. By contrast, we assume that in the inactive state binding costs some energy $\Delta E$, leading to $E_{10}^{In}=E_{01}^{In}=\Delta E$ and $E_{11}^{In}= 2\Delta E$. \textcolor{black}{The last assumption of additivity of  binding energies within a given configuration, i.e. for a frozen mode amplitude, is expected to be accurate if the two binding sites are distant enough, since no elastic coupling between them is expected in that case}.

\begin{figure}
\centering
  \begin{tabular}{@{}c}
    {\includegraphics[width=7.6cm]{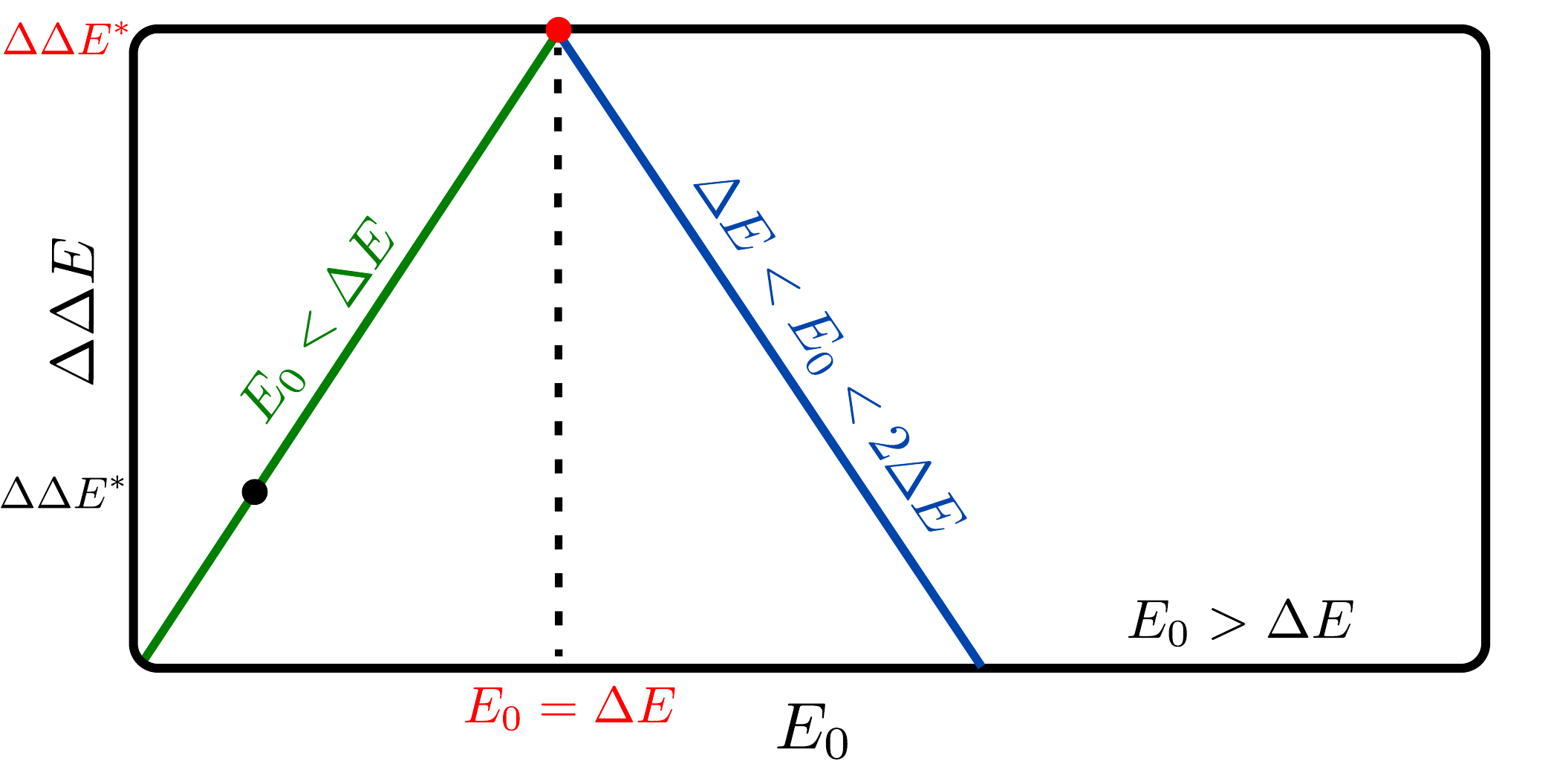}}
	  \end{tabular}
     \caption{Cooperative energy $\Delta\Delta E$ as function of the energy cost of binding a ligand $E_0$ is sketched for the MWC model. The maximal cooperative energy that the system can reach is $\Delta\Delta E^*$. } 
\label{fig:sum}
\end{figure}
 
For each binding situation, the configuration (inactive or active) chosen is the one with the smallest energy, e.g. $E_{01}=\min(E^{In}_{01},E^{Ac}_{01})$. Computing the cooperativity one finds three different cases:

(i) if $E_0<\Delta E$, then the binding of one ligand is sufficient to drive the system in the active state, implying $E_{10}=E_{11}=E_0$ and $\Delta\Delta E=E_0$; 

(ii) if $\Delta E<E_0<2\Delta E$, then the binding of one ligand is not sufficient and two ligands have to bind in order to drive the system in the active state. Consequently, $\Delta\Delta E=2\Delta E-E_0$;

(iii) if $E_0>2\Delta E$, then $\Delta\Delta E=0$ since the system stays in the inactive state even if two ligands are bound.
The sketch of this behavior is shown in Fig.\ref{fig:sum}, illustrating that the maximum cooperativity is found for $E_0=\Delta E$. 

\subsubsection*{Mechanical consideration on the MWC model}
Our observations (see the empirical section below) indicate that a significant fraction of allosteric proteins operate mainly along  one  normal mode
of the elastic energy, supporting the idea that in these cases a favored path connects the inactive and active configurations. We expand the energy (in the absence of ligand) in terms of the motion $x$ along that path (in this two-states case, it is more convenient to chose a coordinate $x$ varying between $-1$ and $+1$ as the protein undergoes its allosteric response). \textcolor{black}{Note that if non-linearities are present, this path is not along a single linear mode, but it bends in phase space. Our analysis below holds independently of such bending. }We keep the minimal number of non-linear terms that allow to display two states (a polynomial of four degrees has five parameters, three of them can be fixed by redefining the reference energy, and changing the definition of $x$ by both a multiplicative and additive constants):
\begin{equation}
\label{eq:phi4}
E_{00}(x)= k_a ||d R_a||^2 \left( \frac{1}{8}x^4 - \frac{1}{4}x^2+ x b\right),
\end{equation}
where $b$ is a parameter reflecting how the energy profile is tilted towards the inactive state,  $k_a$ characterizes the stiffness of the mode and $||d R_a||^2$ is the square norm of the allosteric response. 

\begin{figure}
\centering
  \begin{tabular}{@{}c@{}c}
  \large{\textbf{A}}
   {\includegraphics[width=6.2cm]{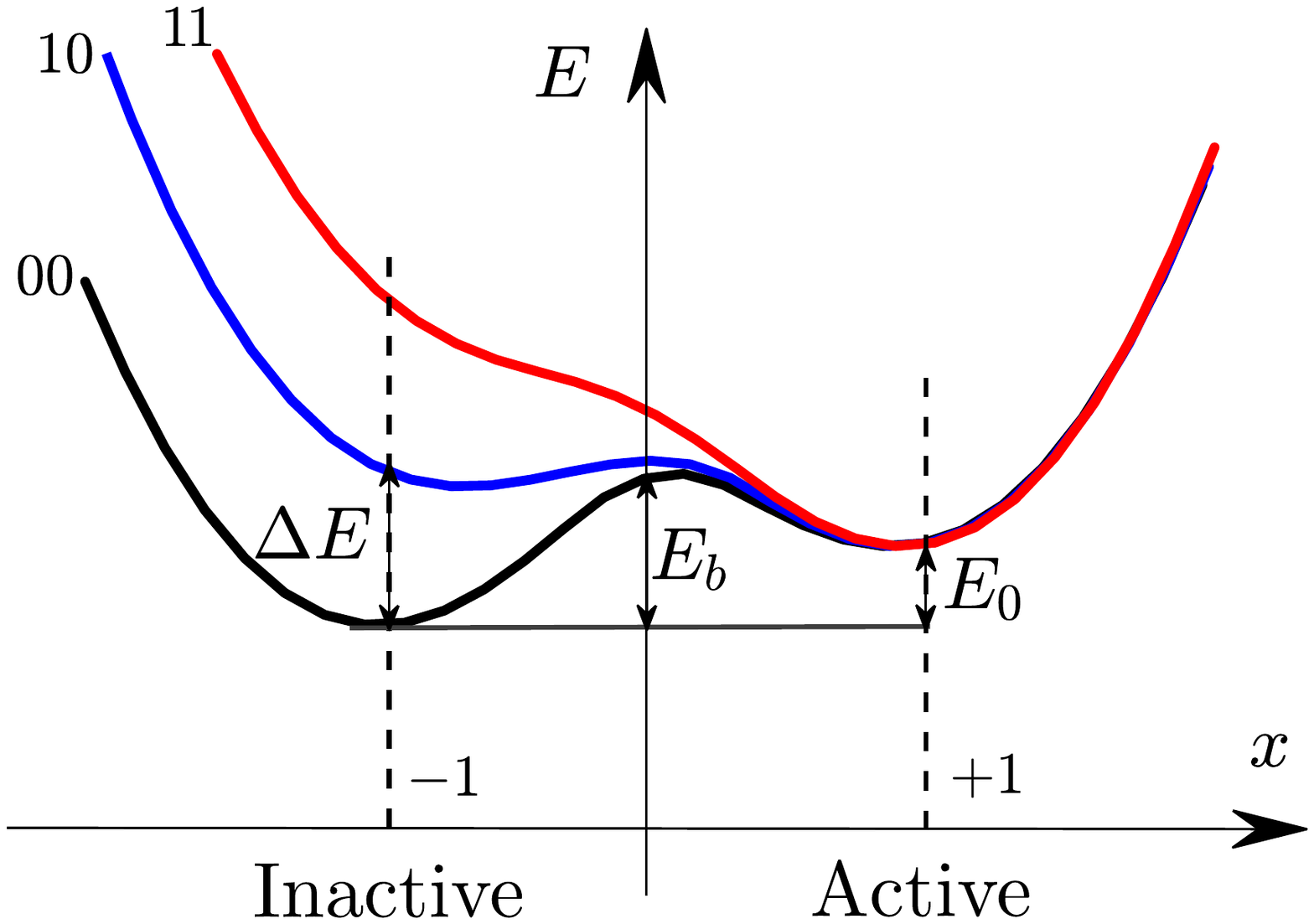}}\\ 
   \large{\textbf{B}}
    {\includegraphics[width=8cm]{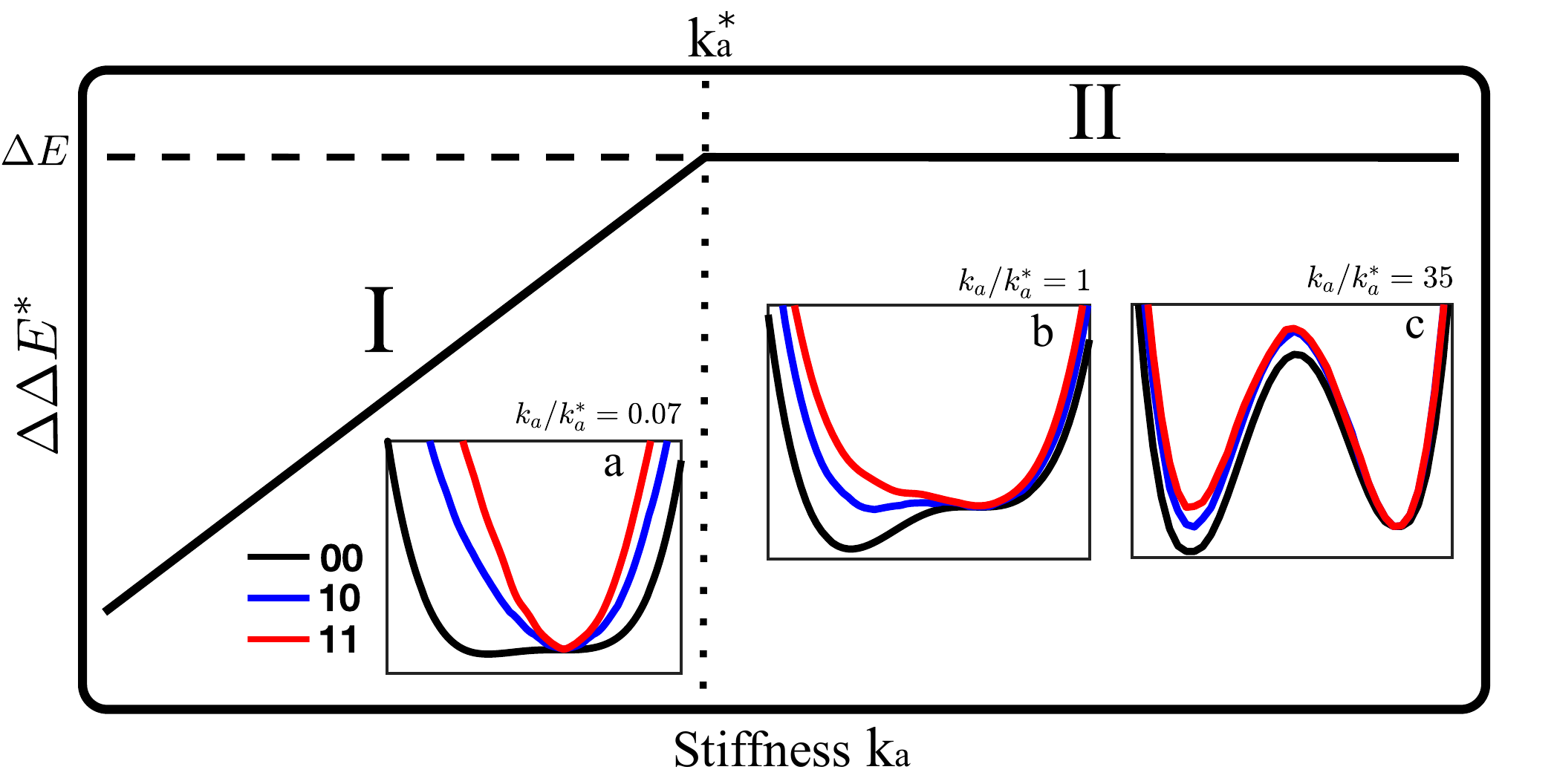}}
	  \end{tabular}
     \caption{(A) Sketch of the energy profiles $E_{00}(x)$ (black), $E_{10}(x)$ (blue) and $E_{11}(x)$ (red) between two states active ($Ac$) and inactive ($In$) as a function of the motion $x$ along the path connecting them. 
   The energy difference $E_0$ between the inactive and active unbound state, the energy cost of binding one ligand $\Delta E$ and the height of the energy barrier $E_b$ are highlighted. (B) Maximal cooperative energy $\Delta\Delta E^*$ as function of the stiffness of the allosteric response $k_a$, showing (I) linear growth and (II) a plateau.  energy profiles for $k_a\ll k_a^*$, $k_a\sim k_a^*$ and $k_a\gg k_a^*$ are show in inset. There are obtained from the non-linear elastic model of allostery described below.  } 
\label{fig:phase}
\end{figure}

A typical profile following Eq.\ref{eq:phi4} is shown in Fig.\ref{fig:phase}A. We denote by inactive the lowest of the two minima and the other one by active. Note that (i) in the case $b=0$, Eq.\ref{eq:phi4} describes two identical minima at $x=\pm 1$ of stiffness $k_a$, separated by an energy barrier $E_b=k_a ||d R_a||^2/8$. (ii) When the parameter $b$ is positive, we have $E_{Ac}>E_{In}$ up to $b=b_c=1/(3\sqrt 3)$ where the active state becomes unstable and only a single stable state is left. 
(iii) At fixed $k_a$, the  energy difference  between the two states $E_0$ is maximal at $b=b_c$, where one finds $E_0=\frac{3}{8}k_a ||d R_a||^2$. (iv) For small $b$, $E_0\simeq 2k_a ||d R_a||^2 b $. In what follows we focus on the case $b<b_c$ where the population shift model lies.

Next, we assume that the active configuration matches the shape of both ligands, so that binding events in that state cost no energy.
However, by moving away from the  configuration the shapes of the protein and ligands do not match anymore and the protein needs to  deform elastically near the binding site. Again assuming that each binding site involves of order $n_0$ atoms, which move by a distance of order $a$ from the active to the inactive state, we have a binding energy $E_{10}(x)-E_{00}(x)= n_0 k a^2 g(x-x_{Ac})$ where $x_{Ac}\approx 1$ is the location of the active state along the path. This energy is exemplified by the difference between the blue and black curves in Fig.\ref{fig:phase}A.  Thus the binding energy in the inactive state follows $\Delta E= n_0 k a^2 g(x_{In}-x_{Ac})\approx 4  n_0 k a^2$, which is independent of the protein size 
and mode stiffness $k_a$. \textcolor{black}{As explained before, if the two binding sites are distant enough, for a given mode amplitude the elastic costs of binding will simply add up: } 
$E_{11}(x)-E_{00}(x)= 2 n_0 k a^2  g(x-x_{Ac})$ as shown in red in Fig.\ref{fig:phase}A.

\subsubsection*{How the mode stiffness  constrains cooperativity}
To quantify this constraint we define the maximal cooperativity over all possible tilts $b$ given $k_a$: $\Delta \Delta E^*\equiv \max_{b} \left\{\Delta\Delta E(k_a,b)\right\}$. We find two regimes:

(i) If $N k_a\ll n_0 k$, then the elastic costs associated with binding are very large compared to $E_0$.
Both $E_{10}(x)$ and $E_{11}(x)$ are peaked close to $x_{Ac}$, as illustrated in \textcolor{black}{Fig.~\ref{fig:phase}B.a}.
Thus $E_0<\Delta E$, implying $\Delta\Delta E=E_0$ according to Fig.~\ref{fig:sum}, which is maximized at $b=b_c$
leading to $\Delta \Delta E = \frac{3}{8}k_a ||d R_a||^2\sim k_a N a^2$. Thus $\Delta \Delta E$ vanishes linearly at small $k_a$, as illustrated in Fig.\ref{fig:phase}B. This result is qualitatively similar to the induced fit case, for which $\Delta \Delta E$ also vanishes linearly as shown in Fig.\ref{fig:sketch_intro}.B.

(ii)  If $N k_a\gg  n_0 k$, then the elastic cost is very small in comparison to $E_0$: $E_{00}(x)$, $E_{10}(x)$ and $E_{11}(x)$ are almost identical as illustrated in \textcolor{black}{Fig.\ref{fig:phase}B.c}. In that regime, cooperativity is optimized by chosing a small tilt fixing  $E_0$ to $\Delta E$ according to Fig.~\ref{fig:sum}, implying $\Delta\Delta E=\Delta E\sim  n_0 k a^2$ which is independent of $k_a$. This  plateau behaviour is represented in Fig.~\ref{fig:phase}B, and appears at $k^*_a\sim n_0 k/N$.

This result represents a fundamental difference with the induced fit case, for which a large stiffness destroys cooperativity.
Indeed in the induced fit scenario a large stiffness implies that the minimal energy is always found for $x\approx 0$ as illustrated in Fig.~\ref{fig:sketch_intro}B.III, implying that binding does not move the protein along that mode, which is thus useless. This state of affairs is ultimately a geometric necessity stemming from the fact that the three curves $E_{00}(x)$, $E_{10}(x)$ and $E_{11}(x)$ must be very close to each other in that regime, and each present  a single minimum. Consequently  the positions of these minima must be very similar in the three cases, leading to $x\approx 0$ independently of binding ligands or not. This geometric necessity vanishes as soon as two minima are present.

Note that although $\Delta\Delta E$ asymptotes to a constant for $k_a\gg k_a^*$, the barrier $E_b$ between the inactive and active states  grows linearly with $k_a$ in that limit. Large barriers would lead to undesirably  slow transition rate between states,  thus we expect that in practice $k_a$ lies reasonably close to $k_a^*$.

\subsubsection*{A mechanical model for population shift allostery}
We seek to model that (i) the allosteric response often takes place mainly along a single vibrational mode.
(ii) Various architectures  can lead to allostery,  including the well-known shear \cite{Mitchell16,Gerstein94} or  hinge \cite{Xu03} designs and others not falling in these categories \cite{Goodey08,mclaughlin12}. Such a  diversity is also found in  {\it in silico} evolution schemes  \cite{Yan18}. Yet, such synthetic architectures always present soft extended regions where most of the strain (i.e. relative motion)  is located. Such  an observation was made in a few proteins \cite{Mitchell16} and will be generalized below. For proteins presenting two stable configurations, we expect these regions to present two possible ways of locally stacking amino acids well. 

\begin{figure}
\centering
  {\includegraphics[width=8.5cm]{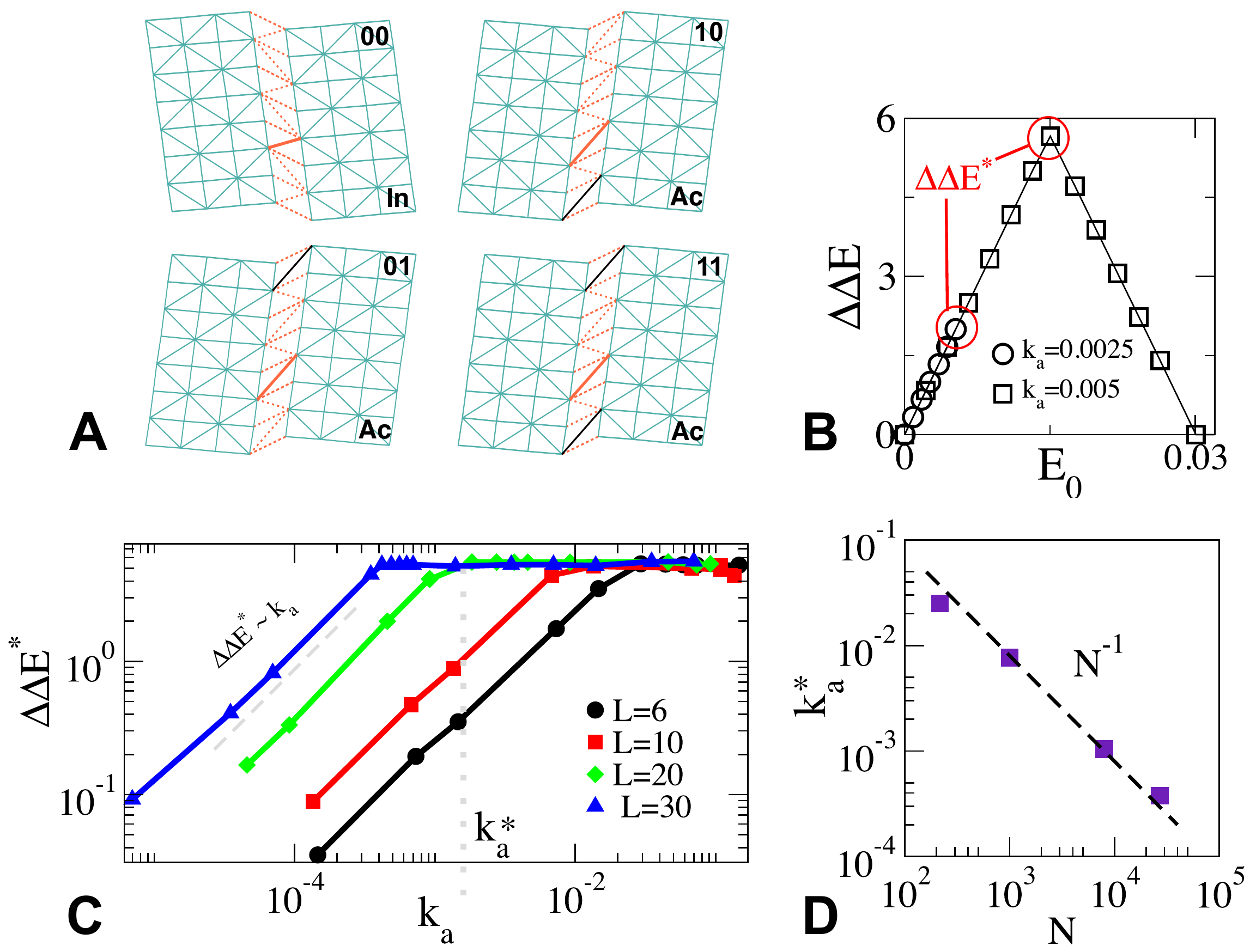}}
	\caption{(A) Sliced view of our three-dimensional mechanical  model for population shift allostery. In blue: rigid elastic regions made of harmonic springs of stiffness $k$, separated by a weak non-linear region made of nonlinear springs (in red) of stiffness $k_w$ with $k_w\ll k$. The inactive state ($In$) favors short springs, while the active state ($Ac$) long springs, as exemplified by a solid orange spring in the weak band. The location of the binding sites are represented by a solid black spring. (B)  $\Delta \Delta E$ as function of $E_0$ for different values of $k_a$ for a linear length $L=20$, where $L^3=N$. (C) Maximal cooperative energy $\Delta \Delta E^*$ as function of  $k_a$ for $L=6,10,20$ and $L=30$. The kink in these curves define the cross-over stiffness $k_a^*$.  (D)  $k_{a}^*$ {\it v.s.} number of nodes in the model $N$, supporting  $k_{a}^*\sim N^{-1}$.}
	\label{fig:numver}
\end{figure}

As an illustration, we consider the shear design in which the protein presents two three-dimensional rigid regions connected by a soft planar layer that can easily deform. The rigid regions consist of  harmonic springs of stiffness $k$, shown in blue in Fig.\ref{fig:numver}A. The soft  layer consists of anharmonic springs (shown in red in Fig.\ref{fig:numver}A), whose energy {\it vs.} extension curve is non-linear, chosen  to have the same form as Eq.\ref{eq:phi4}. These non-linear springs have a characteristic stiffness $k_w$, and present two stable extensions at which they exert no force, whose relative energy is controlled by some bias $b_w$. These stable extensions are chosen such that two states of the protein as a whole, the inactive and active states shown in Fig.\ref{fig:numver}A, present no contact forces by construction, and are thus local minima of the energy. Finally the protein presents two binding sites, at its top and bottom. At each site, binding  imposes that the distance between two nodes (indicated by black lines in Fig.\ref{fig:numver}A)  equals the distance it naturally presents in the active state, thus favoring it. The details about the construction of the microscopic model are discussed in Supplementary Information.

We can now numerically compute the energy profiles $E_{00}(x)$, $E_{01}(x)$ and $E_{11}(x)$ as a function of the motion along the shear mode $x$, by imposing a shear displacement (i.e. a value of $x$) and letting the entire elastic energy of the material relax, except for that mode. The insets of Fig.~\ref{fig:numver} show our results. The value of the mode stiffness $k_a$ can be extracted from fitting Eq.\ref{eq:phi4} to $E_{00}(x)$ and measuring the displacement norm $||d R_a||$, and can be increased by increasing $k_w$. From $E_{00}(x)$ one readily computes the energy difference $E_0$ between the inactive and active states, which can be increased by monitoring the microscopic bias $b_w$. From the minima of $E_{00}(x)$, $E_{01}(x)$ and $E_{11}(x)$ one \textcolor{black}{immediately} extracts the binding costs $\Delta E$ and $\Delta \Delta E$.

Fig.\ref{fig:numver}B shows $\Delta \Delta E$ for two values of $k_a$, as the energy difference $E_0$ is increased.
For large $k_a$, $\Delta \Delta E$ passes through a maximum $\Delta \Delta E^*=\Delta E$, whereas for small $k_a$,  $\Delta \Delta E$ is smaller, and its maximum is fixed by the maximal achievable energy difference $E_0$.  $\Delta \Delta E^*$ as a function of $k_a$ is shown for different system sizes $N$ in Fig.\ref{fig:numver}C, confirming the presence of two regimes with a cross-over at some $k_a^*\sim 1/N$ as shown in Fig.\ref{fig:numver}D. \textcolor{black}{Overall, these observations validate our theoretical predictions on the dependence of $k_a^*$ with the number of residues, and on how the stiffness $k_a$  qualitatively affects cooperativity}.

\subsection*{Empirical study of 34 allosteric proteins} 

\subsubsection*{Allosteric Response}
We identify a set of 34 allosteric proteins in the Protein Data Bank (PDB), for which both the active (ligand bound at the allosteric site) and inactive (no ligand bound at the allosteric site) crystalline X-ray structures are available. Their PDB identifiers are taken from \cite{Daily07, Mitchell16} and reported in Supplementary Information. The set is diverse in functionality, and includes enzymes (13), G-proteins (10), Kinases (3), response regulators (3), DNA-binding proteins (4) and the Human Serum Albumin\textcolor{black}{, among which 12 protein complexes are present}. We can thus estimate the allosteric response $|dR_a\rangle $ as the  displacement field between the inactive and active structures (after having aligned them via the software Pymol 2.1.1 \cite{PyMOL}). Here we focus on the motion of the $N$ amino-acids, located by the position of their $\alpha$-carbon. As an illustration, the allosteric response of a given protein (the elongation factor Tu) is shown in black arrows in Fig.~\ref{fig:all}A.v
From the allosteric response $|dR_a\rangle $, one can readily estimate: (i) the magnitude of the displacement $||dR_a||^2$. 
(ii) The fraction of the protein involved in the response. For any displacement field, this fraction is usually estimated via the participation ratio \cite{Bell70}: 
\begin{equation}
P=\frac{||dR_a||^2}{ N\sum_{i=1}^N ||dR_a(i)||^4}.
\end{equation}
(iii) A measure of how much relative displacement takes place around atom $i$. Following \cite{Mitchell16} we consider the shear pseudo-energy $E_{sh}(i)$ quantifying the amount of strain~---~essentially a measure of the {\it relative displacement} between adjacent atoms~---~at residue $i$, whose precise definition is given in Supplementary Information. $E_{sh}(i)=0$ indicates that the protein moves as a rigid body near atom $i$, and by contrast is large where atoms slide rapidly past each other. $E_{sh}(i)$ is shown in color in Fig.~\ref{fig:all}A for the protein Tu,
illustrating that two parts of the protein are rigidly moving (and counter-rotate), while the central region displays significant pseudo-energy $E_{sh}(i)$, which is reminiscent of a hinge design.

\subsubsection*{Elastic Networks Analysis}
To estimate protein elasticity we use elastic network models (ENM) \cite{Atilgan01}, in which harmonic springs of identical stiffness are placed between all $N$ $\alpha$-carbons laying below a chosen cutoff radius $R_c\in[8-12]$ \r{A}. \textcolor{black}{ENM is obviously a crude  approximation of real atomic interactions. Yet, its simplicity allows for the systematic study of various proteins, and it has been successful in capturing normal modes relevant for the function of some proteins.} The dependence of the results on the value of $R_c$ are indicated by error bars in Fig.~\ref{fig:all}B and discussed in Supplementary Information for Fig.~\ref{fig:all2}B. This procedure defines an elastic energy from which the matrix of the second derivatives, i.e. the Hessian matrix $H$, can be computed. From $H$, one can readily estimate  the stiffness $k_a$ of the allosteric response as the curvature of the elastic energy in that direction:
\begin{equation}
\label{eq:freq}
k_a=\frac{\langle dR_a |H| dR_a\rangle}{|| dR_a||^2}\, ,
\end{equation}
Finally, the eigenvectors of $H$ define the $3N$ vibrational modes of the protein $\{|v_i\rangle\}_{i=1\dots 3N}$.
Following \cite{Kitao99,Bahar99b,Xu03,Rios05,Zheng06}, the overalp $q_i$ between the allosteric response and mode $i$ characterizes their similarity ($q_i=1$ implies that there are identical):
\begin{equation}
q_i=\frac{|\langle dR_a|v_i\rangle|}{\sqrt{\langle dR_a|dR_a\rangle\langle v_i|v_i\rangle}} \, .
\end{equation}
\begin{figure*}
\centering
  \begin{tabular}{@{}c@{}c@{}c}	{\large{\textbf{A}}\includegraphics[width=5.4cm]{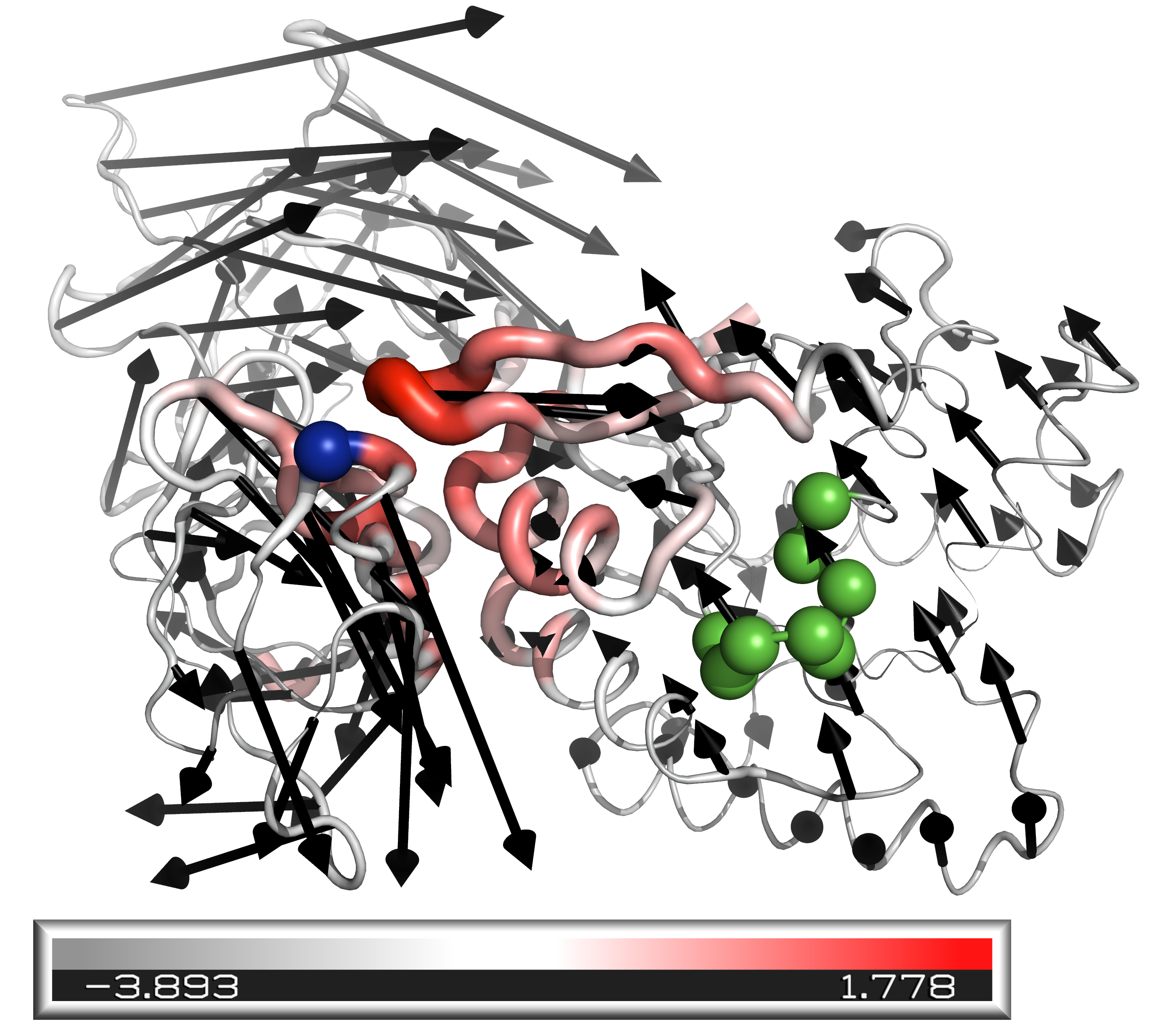}} & {\large{\textbf{B}}\includegraphics[width=5.7cm]{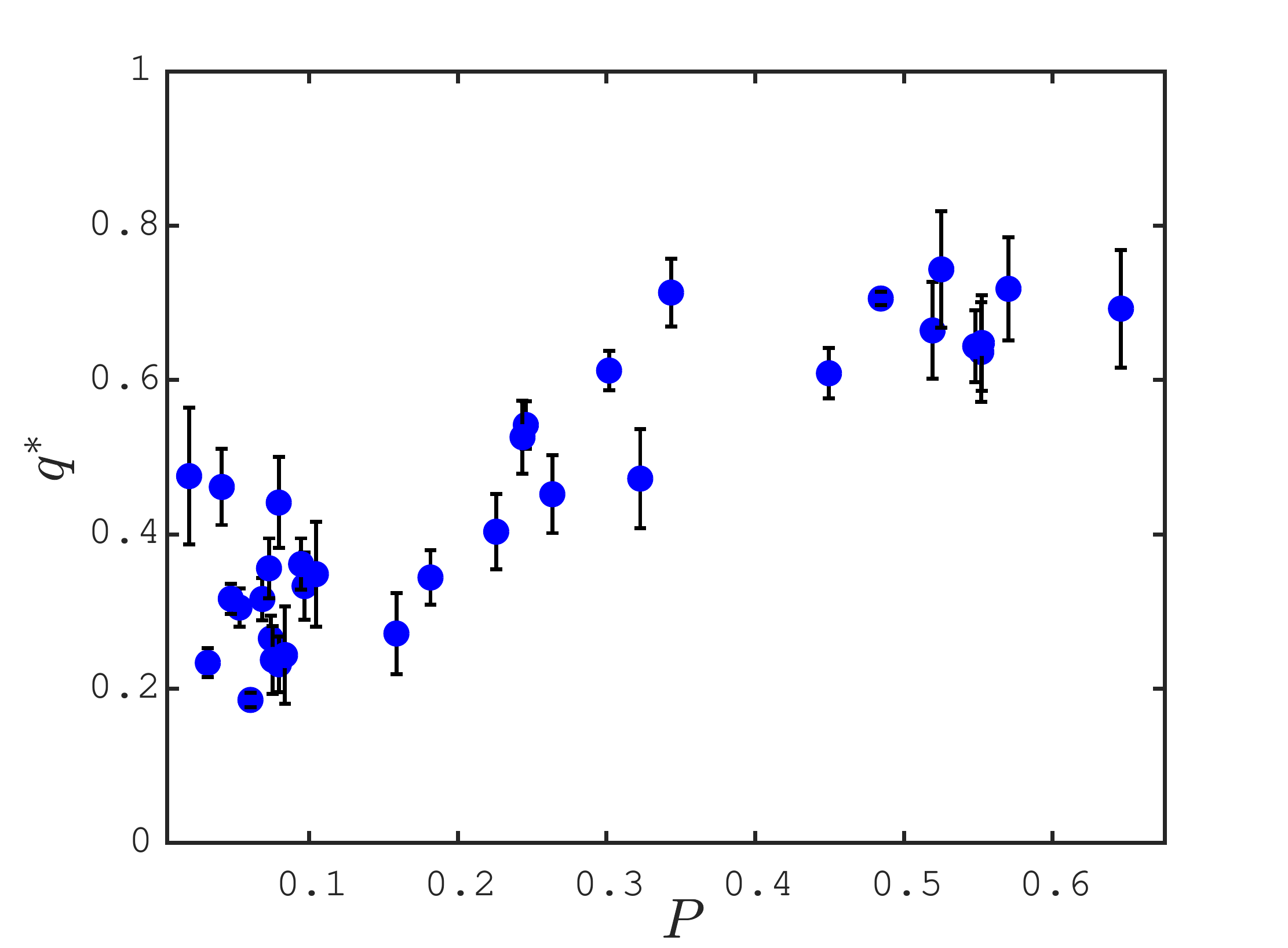}} &
    {\large{\textbf{C}}\includegraphics[width=5.7cm]{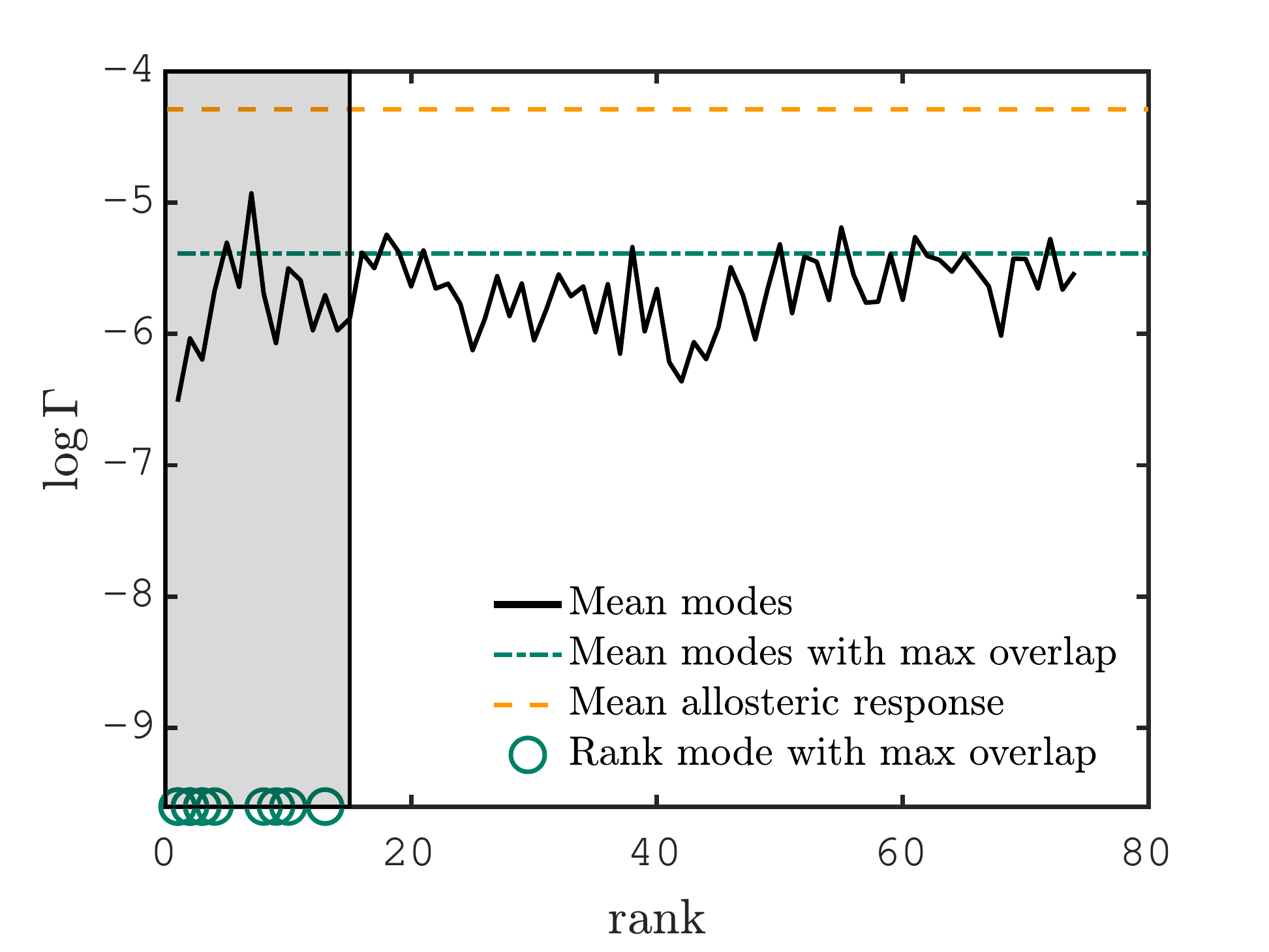}} \\
	  \end{tabular}
	\caption{(A) Allosteric response (black arrows) of elongation factor Tu corresponding to the displacement between the inactive state (where GDP is bound at the allosteric site) and  active state (where GTP is bound at the allosteric site and the aminoacyl-tRNA can bind at the active site). The phosphate binding loop (allosteric site) is highlighted in green, while the active site is at the interface between the GDP binding domain and the two other domains, one residue of which is highlighted in blue \cite{Kjeldgaard93}. The shear is encoded in both the color and the thickness of the structure in a logarithmic scale, red corresponds to large shear. The allosteric response is similar to that of a hinge. (B) Maximal overlap $q^*$ as function of the participation ratio  $P$. (C)  The observable $\log\Gamma$ quantifying how modes are both extended and present large shear energy (see main text for definition) averaged over the proteins with overlap larger than $45\%$, as function of the mode rank for (i) the allosteric response , (ii) the modes with largest overlap  and (iii) the first 75 modes (having subtracted the one with largest overlap) as indicated in legend. The green circles correspond to the rank of the mode with largest overlap. The shaded region highlights the range where these modes fall. }
	\label{fig:all}
\end{figure*}

\subsubsection*{Geometry of the allosteric response}
Fig.~\ref{fig:all}B reports the maximal overlap $q^*\equiv \max_i q_i$, as a function of the participation ratio $P$.  Our observation indicate that $q^*$ is in general large (in half of the case larger than $0.45$),  supporting further that allostery indeed occurs mainly along one mode \cite{Kitao99,Bahar99b,Xu03,Rios05,Zheng06}. Interestingly, this effect is stronger when most sites of the protein are involved in the allosteric response ($P$ large). \textcolor{black}{Interestingly, some protein complexes (for example ATCase, see the values of $q^*$ in Table 1 of Supplementary Material) also display a large projection of their allosteric response on a single mode, supporting that mechanical considerations can be valuable in such cases as well.}

We now provide systematic data supporting that the allosteric response presents extended regions of large shear energy \cite{Mitchell16}. 
More specifically, we argue that while some vibrational modes can present significant shear (e.g. localized modes capturing the motion of dandling loops) and other can be extended (such as plane-wave-like modes), the allosteric response is unique in presenting both aspects, thus revealing a specific design principle.  To quantify this effect, we introduce the quantity, that can be defined on any displacement field:
\begin{equation}
    \log\Gamma\equiv \left[\gamma\log_{10}(\textrm{P})+\log_{10}(||E_{sh}||)\right]
\end{equation}
where $||E_{sh}||$ is the total magnitude of the shear energy, i.e. $||E_{sh}||=(\sum_i E_{sh}(i)^2)^{1/2}$.
$\log\Gamma$ is large if the displacement is extended and if the shear energy is large. The factor $\gamma$ characterizes the trade-off between these two features. Here we choose  $\gamma=3.5$ reflecting the fact that for vibrational modes, we find that $P$ varies about 3.5 times less in relative terms than $||E_{sh}||$ as shown in Supplementary Information. Thus for $\gamma=3.5$, the spatial extension and the amount of shear equally affect $\log\Gamma$. 
 Fig.~\ref{fig:all}C shows  $\log\Gamma$  averaged over the $17$ proteins with $q^*>45\%$ for  the allosteric response (yellow line), the mode with maximum overlap (blue line) and the first $75$ vibrational modes (having subtracted the one with largest overlap) as function of the mode rank. We find that $\Gamma$ is typically 160 times larger for the allosteric response than for vibrational modes, a very significant difference underlying the specific geometry of the allosteric response.

\subsubsection*{Scaling of response stiffness $k_a$ with protein size}
\begin{figure*}
\centering
  \begin{tabular}{@{}c@{}c@{}c}
  {\large{\textbf{A}}\includegraphics[width=6cm]{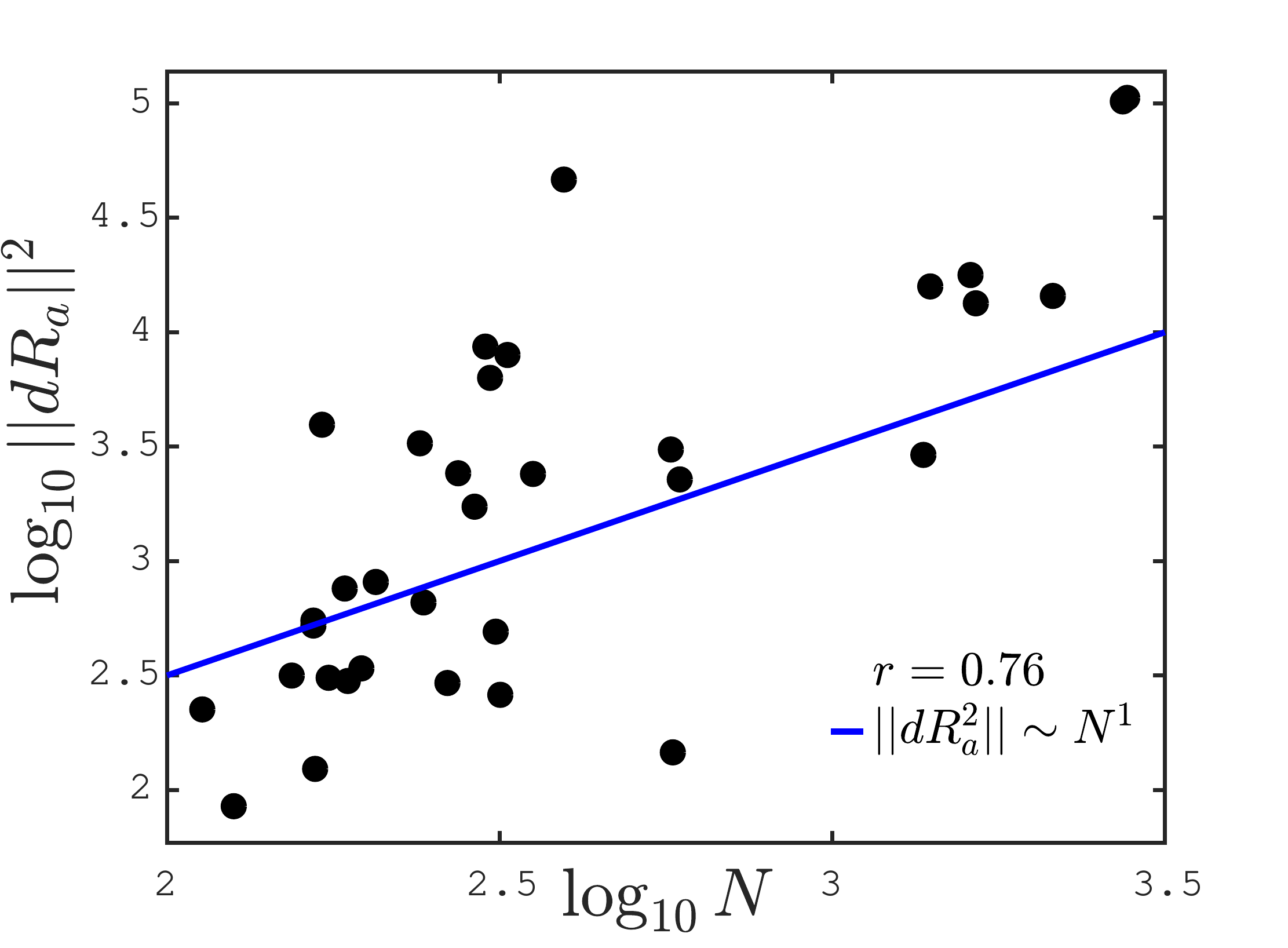}}&
  {\large{\textbf{B}}\includegraphics[width=6cm]{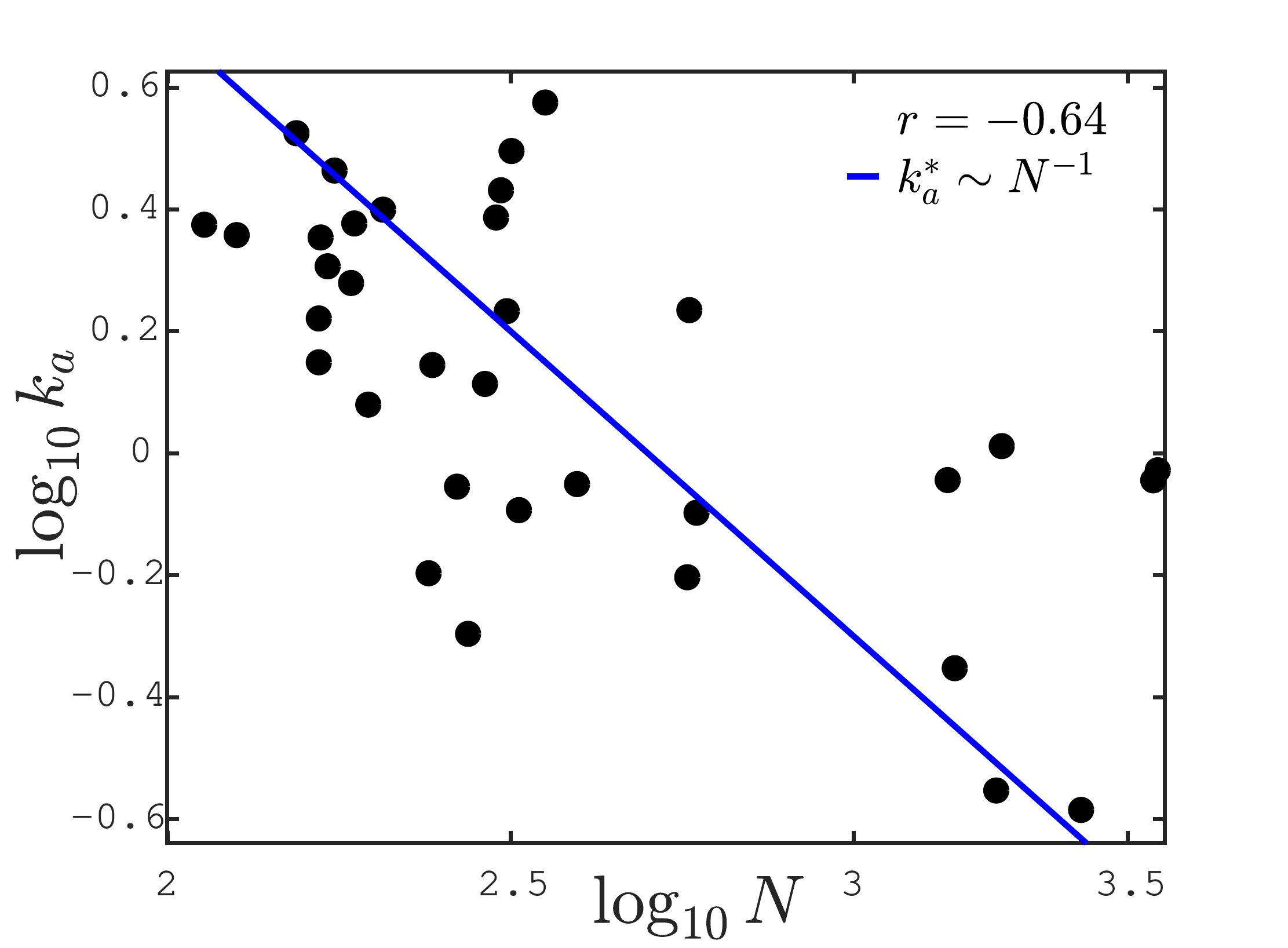}}\\
	  \end{tabular}
	\caption{
	(A) Logarithm of the square of the norm of the allosteric response, $||dR_a||^2$, is shown as function of the logarithm of the number of residues, $N$. The solid line corresponds to $||dR_a^2||\sim N$. 
	(B) The logarithm of the stiffness $k_a$ as function of the logarithm of the number of residues  $N$.  The solid line represents the theoretical prediction  $k_{a}^*\sim 1/N$.  In both plots $r$ indicates the Pearson coefficient.}
	\label{fig:all2}
\end{figure*}

We can now test our conjecture that the allosteric stiffness $k_a$ is close to $k_a^*\sim 1/N$ where cooperativity saturates, which implies in particular an anti-correlation between $k_a$  and protein size.
In our theoretical estimate  of $k_a^*$, we have assumed that the allosteric response magnitude is linear in the protein size, i.e. $||d R_a||^2\sim N$. \textcolor{black}{It is a natural assumption since the larger the protein, the more likely its response involves many residues.} The relationship between these two quantities is tested in 34 proteins in Fig.\ref{fig:all2}A. We indeed find a strong correlation between the logarithms of $||d R_a||$ and $N$  (Pearson coefficient $r=0.76$ \textcolor{black}{with p-value $p=1.64 \times 10^{-7}$. Pearson coefficients are computed on the logarithmic values via the Matlab R2017b function \textit{corr}}). Overall, data are consistent with our assumption of proportionality. 

Finally, we plot the allosteric response stiffness $k_a$ measured according to Eq.\ref{eq:freq} in terms of $N$ for all proteins in Fig.~\ref{fig:all2}B. A key finding is the fair anti-correlation between the logarithms of these two quantities (Pearson coefficient $r=-0.64$ \textcolor{black}{with p-value $p=4.00\times 10^{-5}$}), supporting the idea that larger allosteric proteins need to evolve a softer elastic mode to accomplish function, as expected from our analysis. \textcolor{black}{The signal that we observe is encouraging, especially given the possible sources of noise in the analysis, among which, in particular, the use of ENM to build the Hessian matrix.}

\section*{Conclusion}
We have provided systematic evidence that the allosteric response occurs along one soft elastic mode, and we have introduced a novel observable $\Gamma$ to establish that this response generally displays unusually  extended regions of high shear strain. These observations support that for many proteins,  elasticity is a useful starting point to describe allostery. We have revisited the two classical thermodynamic models of allostery from this perspective, and provided a detailed study of how the energy profile along the soft mode evolves with binding.
We find that induced fit and population shift models qualitatively differ. In the induced fit model, there is an optimal stiffness $k_a^*\sim 1/N$ associated to that mode beyond which the cooperative binding energy eventually decreases to zero. The population shift model is more robust to mutations affecting stiffness, and $k_a^*\sim 1/N$ simply marks a cross-over beyond which  cooperativity saturates and the transition time between configurations rapidly explodes. We introduced a novel non-linear elastic model for allostery supporting these views. Our key  result is that proper function is achieved if proteins evolve an elastic mode whose softness must rapidly decrease with size, a prediction  supported by the  anti-correlation observed between these quantities.

Systematic mutation scan on one single protein, in which binding assays to measure cooperativity are combined with single molecule experiments or ultrafast laser pulses  to estimate the stiffness of the allosteric response, would be extremely useful to test the predicted relationship between these  quantities.  Molecular dynamics experiments could further test how the energy profile along the soft elastic mode evolves with binding. Elucidating such an interplay between thermodynamics and mechanics in proteins would be valuable in a variety of tasks, including {\it de novo} protein design and the discovery of novel allosteric pathways.

\begin{acknowledgements}
We thank P. Barth, B. Bravi, P. De Los Rios, D. Malinverni, R. B. Phillips and M. Popovi\'c for discussions. R.R. is supported by the Swiss National Science Foundation under Grant No. 200021-165509/1. M.W. thanks the Swiss National Science Foundation (Grant No. 200021-165509) and the Simons Foundation (Grant \#454953 Matthieu Wyart). This material is based upon work performed using computational resources supported by the ``Center for Scientific Computing at UCSB'' and NSF Grant CNS-0960316, by the ``High Performance Computing at NY'' and by the ``High Performance Computing at EPFL''.
\end{acknowledgements}

%

\clearpage

\section{Supplemental Material}

\setcounter{equation}{0}
\setcounter{figure}{0}
\makeatletter 
\renewcommand{\theequation}{S\arabic{equation}}
\renewcommand{\thefigure}{S\arabic{figure}}

\section*{A. Mechanics of allosteric proteins from X-ray structural data}
\subsection*{Dataset}
We consider 34 allosteric proteins with both inactive and active X-ray structures available \cite{Daily07, Mitchell16}. The Protein Data Bank (PDB) entries of these proteins are listed in Table \ref{tab:1}; the number of common residues between the inactive and active structures and the number of chains that we consider are also reported in the table. We remove the first two and last two residues for every structure as they correspond to the fluctuating starting and ending points of the protein chain.

\begin{table*}[h]\centering
\caption{Table with the X-ray structures (PDB identifier) used in the analysis. The number of the common residues and of the considered chains are also reported, along with the participation ratio of the allosteric response and the maximal overlap found between the allosteric response and the normal modes. The PDB identifiers are taken from \cite{Daily07} and \cite{Mitchell16} (asterisk).}
\begin{tabular}{lrrrrrr}
Protein type & Inactive (PDB) & Active (PDB) & Common residues & Chains & Overlap & Participation ratio\\
\midrule
1. \, arf6 & 1E0S & 2J5X & 160 & A & 0.27 & 0.16\\
2. \, cdc42 & 1AN0 & 1NF3 & 183 & AB & 0.33 &0.09\\
3. \, rab11 & 1OIV & 1OIW & 162 & A & 0.32 & 0.07\\
4. \, rac1 & 1HH4 & 1MH1 & 175 & A & 0.27 &0.07\\
5. \, ras & 4Q21 & 6Q21 & 164 & A & 0.44 & 0.08\\
6. \, rheb & 1XTQ & 1XTS & 165 & A & 0.35 & 0.1\\
7. \, rhoA & 1FTN & 1A2B & 173 & A & 0.24 & 0.08\\
8. \, sec4 &  1G16 & 1G17 & 152 & A & 0.23 & 0.08\\
9. \, IGF-1R & 1P40 & 1K3A & 283 & A & 0.23 & 0.03\\
10. met repressor  & 1CMB & 1CMA & 204 & AB & 0.24 & 0.08\\
11. tet repressor & 2TRT & 1QPI & 190 & A & 0.47 & 0.32\\
12. glcN-6-P deaminase & 1CD5 & 1HOT & 262 & A & 0.45 & 0.26\\
13. EF-Tu & 1TUI & 1EFT & 393 & A & 0.61 & 0.30\\
14. $\rm{G}_{\rm{t}\alpha}$ & 1TAG & 1TND & 310 & A & 0.36 & 0.07\\
15. ERK2 & 1ERK & 2ERK & 347 & A & 0.19 & 0.06\\
16. IRK & 1IRK & 1IR3 & 296 & A & 0.31 & 0.05\\
17. lac repressor & 1TLF & 1EFA & 536 & AB & 0.71 & 0.48\\
18. PurR & 1DBQ & 1WET & 271 & A & 0.71 & 0.34\\
19. anthranilate synthase & 1I7S & 1I7Q & 1402 & ABCD & 0.61 & 0.45\\
20. chorismate mutase & 2CSM & 1CSM & 241 & A & 0.54 & 0.25\\
21. FBPase-1 & 1EYJ & 1EYI & 323 & A & 0.46 & 0.04\\
22. phosphofructokinase & 6PFK & 4PFK & 315 & A & 0.32 & 0.05\\
23. PTB1B & 1T48 & 1PTY & 287 & A & 0.48 & 0.02\\
24. ATCase$^*$ & 6AT1 & 8AT1 & 2732 & ABCDEFGHIJKL & 0.64 & 0.55\\
25. hemoglobin & 4HHB & 1HHO & 570 & ACBD & 0.66 & 0.52\\
26. NAD-malic enzyme & 1QR6 & 1PJ2 & 2144 & ABCD & 0.72 & 0.57\\
27. phosphoglycerate DH & 1PSD & 1YBA & 1580 & ABCD & 0.74 & 0.53\\
28. human serum albinum$^*$ & 1E78 & 2BXB & 574 & A & 0.36 & 0.09\\
29. fixJ & 1DBW & 1D5W & 238 & AB & 0.69 & 0.65\\
30. DAHP synthase & 1KFL & 1N8F & 1340 & ABCD & 0.64 & 0.55\\
31. SpoIIAA & 1H4Y & 1H4X & 106 & A & 0.40 & 0.22\\
32. CheY & 3CHY & 1FQW & 124 & A & 0.34 & 0.18\\
33. glycogen phosphorylase & 1GPB & 7GPB & 1626 & AB & 0.53 & 0.24\\
34. ATCase & 1RAC & 1D09 & 2774 & ABCDEFGHIJKL & 0.65 & 0.55\\
\bottomrule
\end{tabular}
\label{tab:1}
\end{table*}

\subsection*{Elastic network model}
From the X-ray structures we compute the Hessian matrix $H$ using the anisotropic network model introduced in \cite{Atilgan01}. We consider the positions of residues by looking only at {alpha-carbon $C_{\alpha}$ atoms; ${\bf R}_i$ for alpha-carbon $i$}. The model assumes a harmonic interaction between two residues at distance smaller than a cutoff distance $R_c$, 
\begin{equation}
    V_{ij}=\frac{k}{2}(l_{ij}-l_{ij}^0)^2\, ,
\end{equation}
where $k$ is the spring constant (fixed to unity), {$l_{ij}=||{\bf R}_i-{\bf R}_j||$} is the distance between residues $i$ and $j$, and $l_{ij}^0<R_c$ is the distance at equilibrium. Building the Hessian is then straightforward by taking second derivatives of the potential $V_{ij}$ with respect to the coordinates of residues and  evaluated at equilibrium $l_{ij}=l_{ij}^0$, {
\begin{equation}
    H_{i\mu,j\nu} = \sum_{kl}\left.\frac{\partial^2 V_{kl}}{\partial R_{i\mu}\partial R_{j\nu}}\right|_{l_{kl}=l_{kl}^0}
    = \frac{k}{2}\sum_{kl}\frac{\partial l_{kl}}{\partial R_{i\mu}}\frac{\partial l_{kl}}{\partial R_{j\nu}},
\end{equation}
where $\mu,\nu=1,2,3$ label the spatial dimension of the atoms.
}

\begin{figure*}[h]
\centering
\begin{tabular}{@{}c@{}c@{}c}
  {\large{\textbf{A}}\includegraphics[width=8cm]{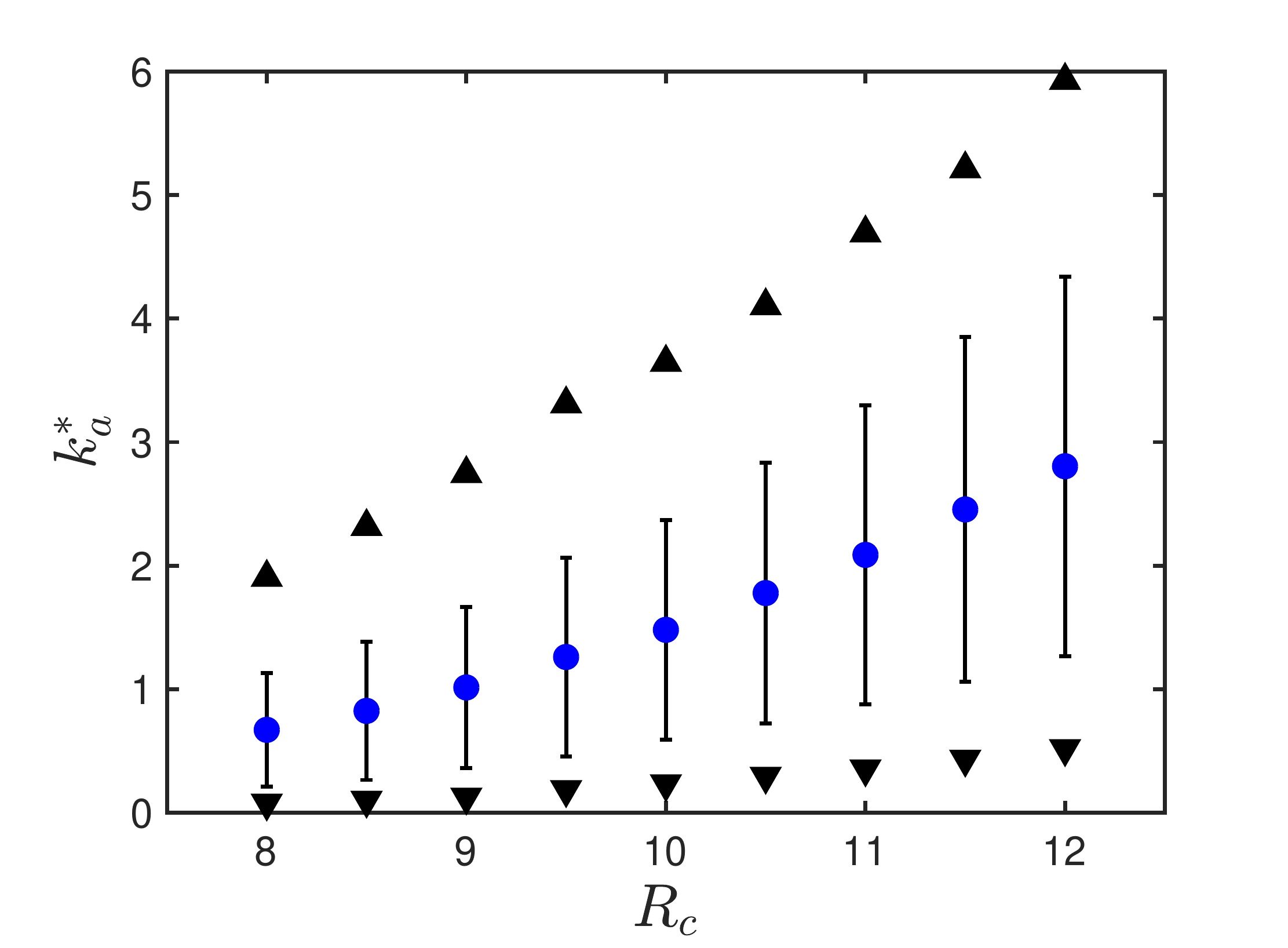}}&
  {\large{\textbf{B}}\includegraphics[width=8cm]{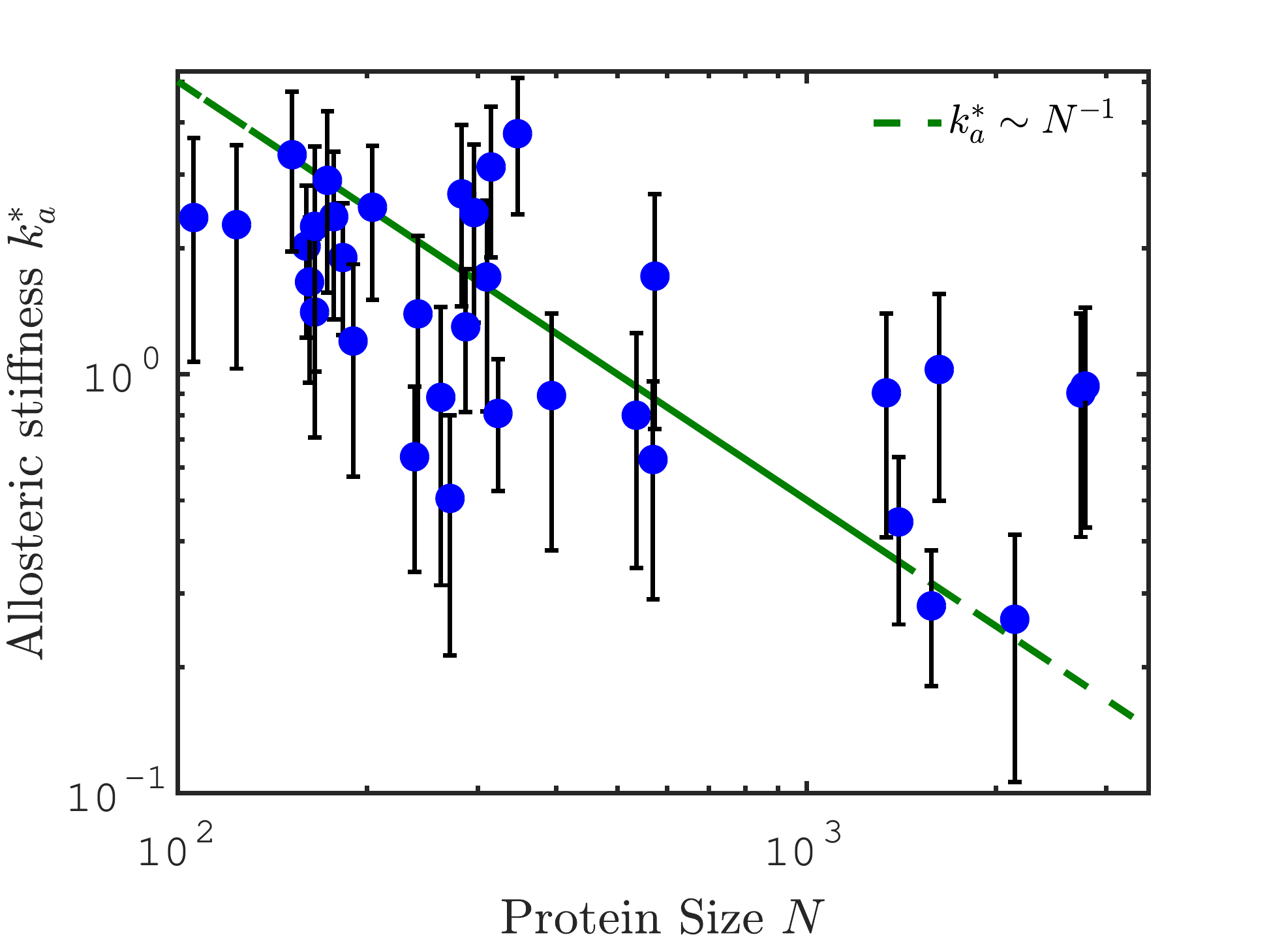}}\\
\end{tabular}
\caption{(A) We plot the stiffness of the allosteric response $k_a*$ as function of $R_c$, where we choose nine equidistributed values of $R_c$ in the range $[8-12]$ \r{A}. We show the mean of $k_a^*$ over all 34 proteins for each $R_c$ (blue circles) with vertical bars given by the standard deviation. We also show the values of $k_a^*$ versus $R_c$ for the protein with largest (upward triangle) and smallest (downward triangle) values of $k_a^*$. Increasing the cutoff distance systematically increases the stiffness of all the points. (B) The stiffness associated to the allosteric response for the data set of allosteric proteins is shown as function of the number of residues in each couple, $N$. The vertical bars represent the standard deviation associated to the distribution of $k_a^*$ resulted from the different values of $R_c$ considered for each protein.}
\label{fig:Variability}
\end{figure*}

The definition of a cutoff distance is an empirical fitting parameter. In practice, we computed Hessian matrices at nine cutoff values equi-distributed in the range $[8-12]$ \r{A}. We define the stiffness of the allosteric response $|dR_a\rangle $ as
\begin{equation}
\label{eq:freq}
k_a^*=\frac{\langle dR_a |H| dR_a\rangle}{\langle dR_a|dR_a\rangle}\, .
\end{equation}
This quantity changes systematically with $R_c$ as shown in Fig.\ref{fig:Variability}A: by increasing $R_c$, $k_a^*$ also increases, which leads to all points in Fig. \ref{fig:Variability}B moving  together in the same direction when changing $R_c$. In the main text we reported the mean value of $k_a^*$ over the nine choices of $R_c$, and in Fig. \ref{fig:Variability}B we show the range of variability of $k_a^*$ when $R_c\in[8-12]$ \r{A}.

\subsection*{Computing the local strain tensor in a protein}
In a continuous medium, a motion maps a point $\vec{X}$ in the reference configuration to a new point ${\vec{x}}$ in the current configuration, the strain tensor of the motion can thus be computed as
\begin{equation}
{\epsilon}_{ab}(\vec{X})=\frac{1}{2}\lp\frac{\partial\vec{x}}{\partial X_a}\cdot\frac{\partial\vec{x}}{\partial X_b}-\delta_{ab}\rp,
\end{equation}
where $a$, $b$ labels the spatial dimension. 

In a discrete medium like proteins which are a collection of atoms (or residues as we consider), computing the partial derivative $\overset\leftrightarrow{\Lambda}=\partial\vec{x}/\partial\vec{X}$ at residue $i$ is not straightforward. Ideally, for any neighboring residue $j$ close enough in space
\begin{equation}
\Delta \vec{x}_{ij}=\overset\leftrightarrow{\Lambda}_i\cdot\Delta\vec{X}_{ij},
\label{eq_lamb}
\end{equation}
where $\Delta\vec{X}_{ij}=\vec{R}_{i0}-\vec{R}_{j0}$ and $\Delta\vec{x}_{ij}=\vec{R}_i-\vec{R}_j$ in our setting, where $\vec{R}_i$ is the position of residue $i$ taken from the X-ray structure. 
We have $n_b$ number of such equations for $\overset\leftrightarrow{\Lambda}_i$ when $n_b$ neighbors are considered. 
So $\overset\leftrightarrow{\Lambda}_i$ are usually over-determined when we consider all neighbors below a certain cutoff distance $R_c$ (we choose $R_{c1}=8.5$ \r{A} for first nearest neighbors and $R_{c2}=10.5$ \r{A} for second nearest neighbors). 
Instead of solving Eq. (\ref{eq_lamb}), we define a mean squared error function \cite{Gullett07}
\begin{equation}
MSE(i)=\sum_{j}(\Delta \vec{x}_{ij}-\overset\leftrightarrow{\Lambda}_i\cdot\Delta\vec{X}_{ij})^2w_j(i),
\end{equation}
where we have kept a weight function $w_j(i)$ of node $j$ contribution to $i$ in general. Specifically, we set as in \cite{Mitchell16} $w_j(i)=1$ for all nearest neighbors to $i$ ($R_{ij}<R_{c1})$, $w_j(i)=1-\dfrac{R_{ij}-R_{c1}}{R_{c2}-R_{c1}}$ for $R_{c1}<R_{ij}<R_{c2}$ and $w_j=0$ otherwise. By minimizing the mean squared error with respect to $\overset\leftrightarrow{\Lambda}_i$, we have
\begin{equation}
\overset\leftrightarrow{\Lambda}_i=\sum_j\Delta\vec{x}_{ij}\Delta\vec{X}_{ij}w_j(i)\cdot\lp\sum_j\Delta\vec{X}_{ij}\Delta\vec{X}_{ij}w_j(i)\rp^{-1},
\end{equation}
and 
\begin{equation}
\overset\leftrightarrow{\epsilon}(i)=\frac{1}{2}\lp\overset\leftrightarrow{\Lambda}_i^t\cdot\overset\leftrightarrow{\Lambda}_i-\overset\leftrightarrow{\delta}\rp,
\end{equation}
where $\overset\leftrightarrow{\delta}$ is the identity tensor.

The shear pseudo-energy \cite{Mitchell16}, a vector field whose components contain a measure of the relative motion of each residue, can be defined from the strain tensor $\overset\leftrightarrow{\epsilon}(i)$ computed above 
\begin{equation*}
E_{sh}(i)=\frac{1}{2}\sum_{l,m=1}^3[\gamma_{lm}(i)]^2\, ,
\end{equation*} 
where the local shear tensor $\overset\leftrightarrow{\gamma}(i)=\overset\leftrightarrow{\epsilon}(i)-(1/3) \text{Tr}[\overset\leftrightarrow{\epsilon}(i)]\overset\leftrightarrow{\delta}$ depends only on the displacement between the two conformations via the strain tensor $\overset\leftrightarrow{\epsilon}(i)$ and $\overset\leftrightarrow{\delta}$.

\subsection*{Empirical definition of $\Gamma$}
\begin{figure*}
\centering
  \begin{tabular}{@{}c@{}c}	{\large{\textbf{A}}\includegraphics[width=8cm]{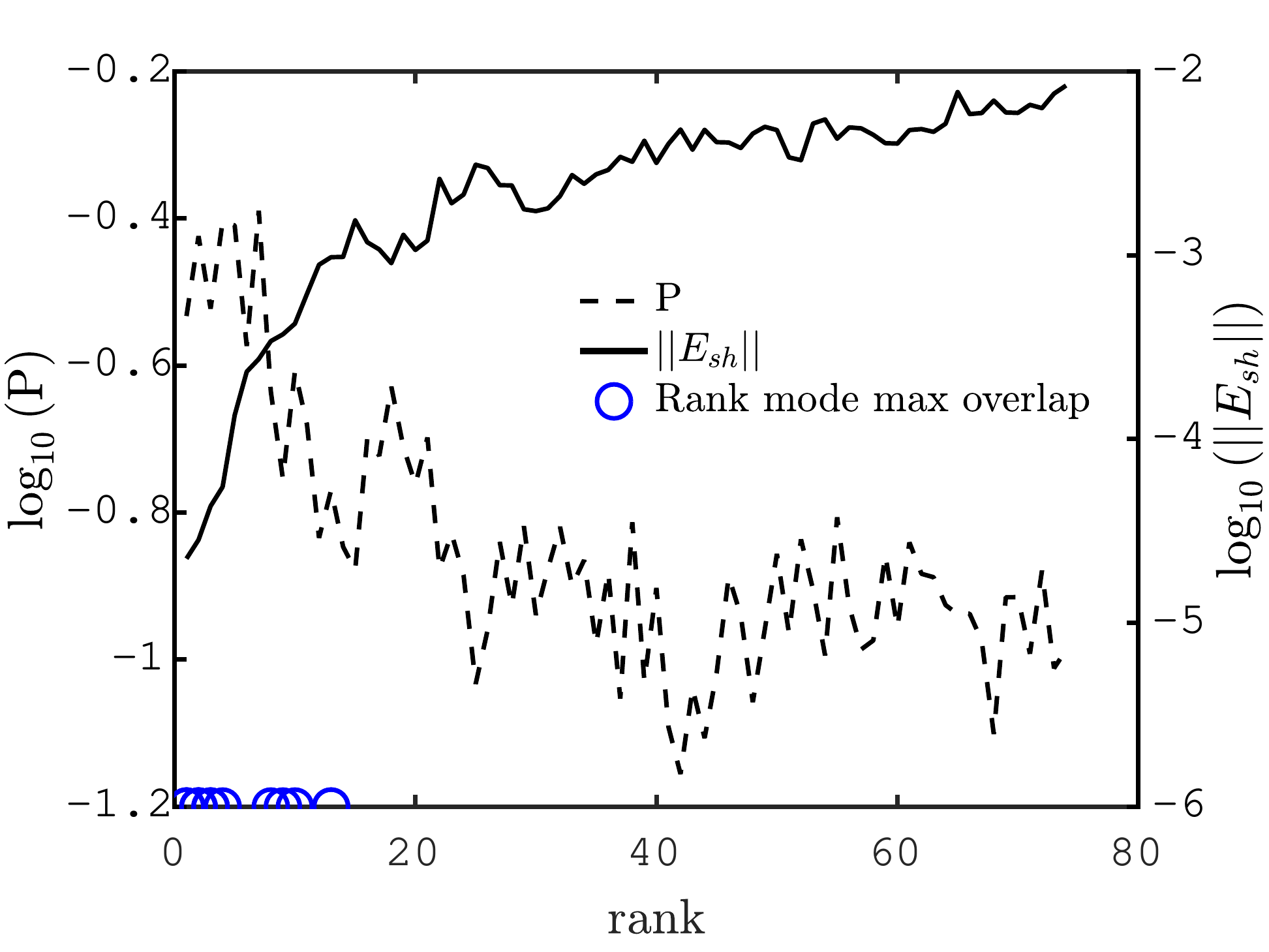}} & {\large{\textbf{B}}\includegraphics[width=8cm]{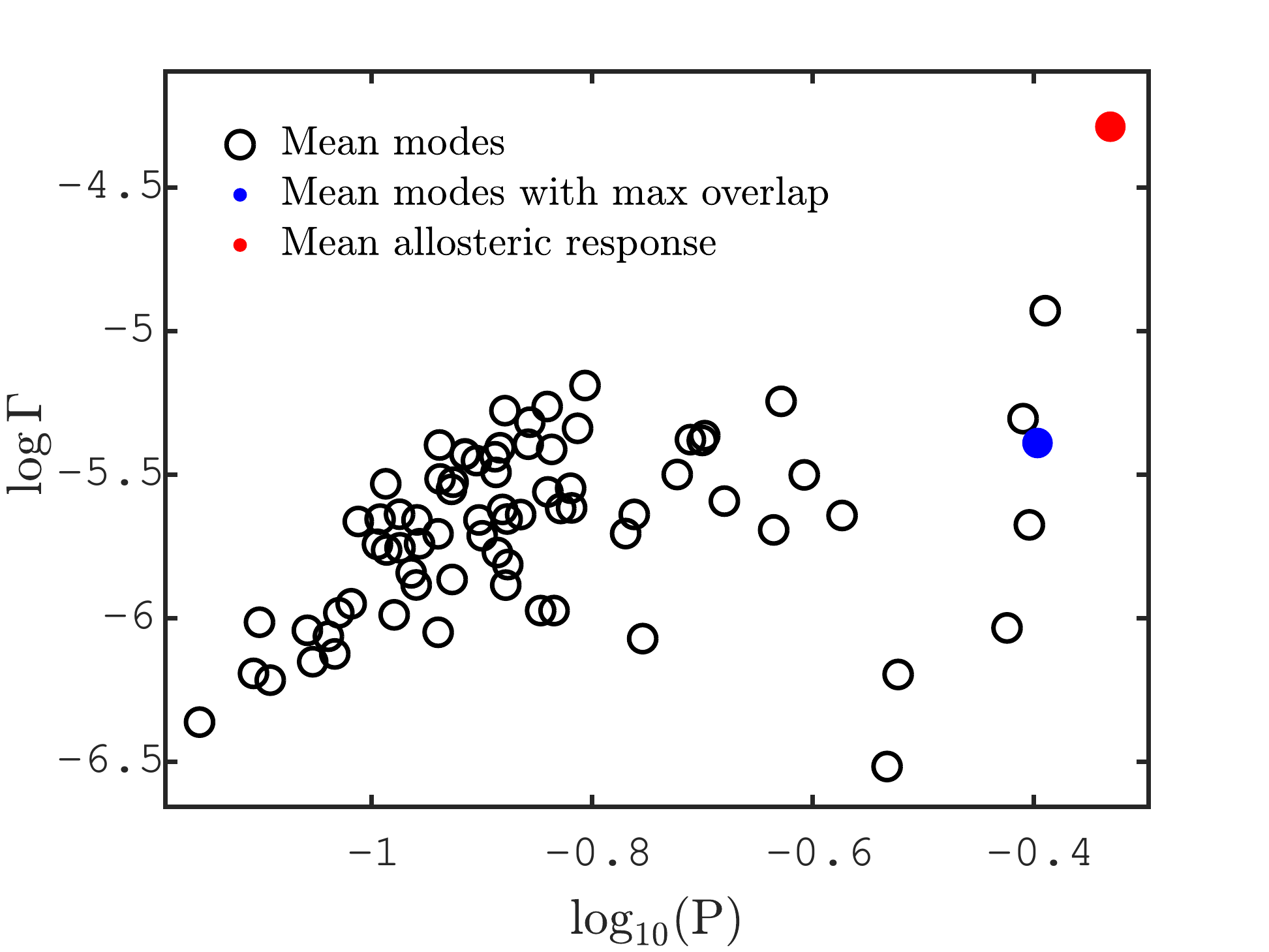}}	  \end{tabular}
    \caption{(A) The logarithm of the participation ratio P and of the norm of the shear pseudo-energy $||E_{sh}||$ is shown as function of the rank. The measured relative variation from the two curves is $\gamma^*\simeq 3.46$ and we choose $\gamma=3.5$ in Eq.\ref{eq:Gamma}. (B) Scatter plot of $\Gamma$ versus the participation ratio P where $\Gamma$ is averaged over the proteins with overlap larger than $0.45$ and shown for (i) the modes (without considering the modes with maximal overlap), (ii) the modes with maximum overlap and (iii) the allosteric response. The different points correspond to different values of mode rank. This alternative visualization confirms the results presented in the main text. }
	\label{fig:Gamma}
\end{figure*}

In the main text we introduced the observable $\Gamma$ combining information on the participation ratio (P) and the amount of shear pseudo-energy ($||E_{sh}||$) in a given mode as
\begin{equation}
\label{eq:Gamma}
    \log\Gamma\equiv \left[\gamma\log_{10}(\textrm{P})+\log_{10}(||E_{sh}||)\right]
\end{equation}
with $\gamma=3.5$. In Fig. \ref{fig:Gamma}.A we show the range of values of participation ratio (dashed line) compared with the one of shear pseudo-energy (continuous line). 

{The measured relative variation from the two curves is $\gamma^*\simeq3.46$ and we choose $\gamma=3.5$ in the definition of $\Gamma$. Fig. \ref{fig:Gamma}.B is an alternative way of representing the result discussed in the main text. We consider a scatter plot of $\log\Gamma$ and participation ratio for the three quantities considered (modes with maximum overlap, other modes and allosteric response) and we see that the allosteric response although displaying a large participation ratio it has also a large $\Gamma$, meaning that it involves also a considerable amount of shear.

\section*{B. Microscopic model}

To construct a microscopic model with two global minima that correspond to the  ``Inactive'' ($In$) and ``Active'' ($Ac$) states, we need to {impose that} all nonlinear springs to stay exactly at one of their minimum when the network is in either $In$ or $Ac$ state. 

\begin{figure*}
\centering
\includegraphics[width=14.8cm]{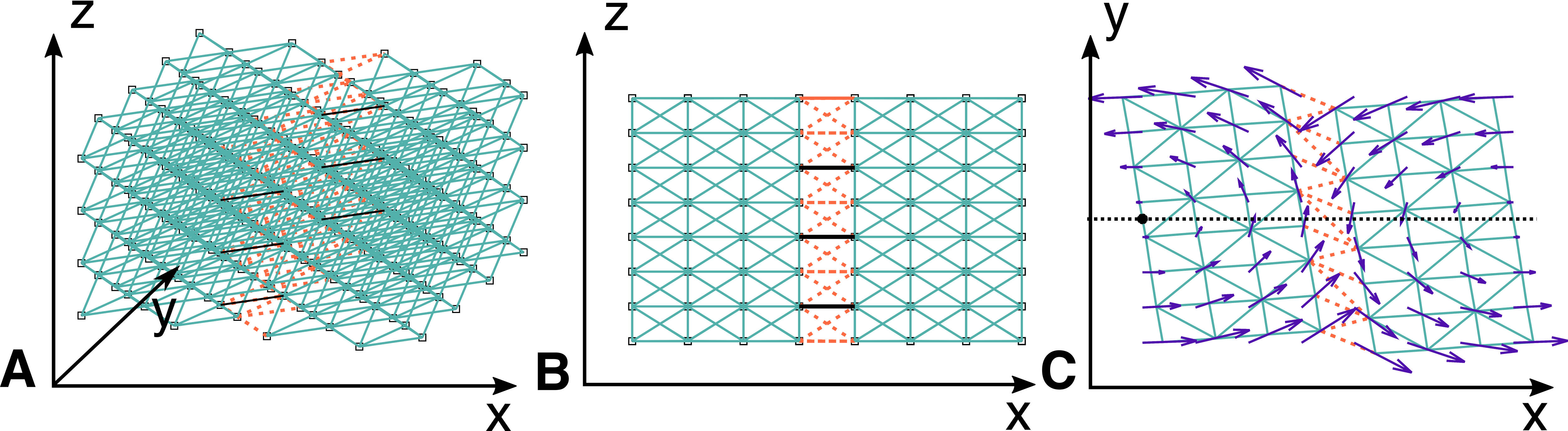}
\caption{(A) Face-centered cubic lattice with open boundaries.  Harmonic springs of stiffness $k$ are represented in clear blue and non-linear springs by  red dashed lines. Black lines represent the links with stiffness $k_1\gg k$, used to simulate the binding of a ligand. (B) A projection of the cubic lattice in the (x,z) plane. (C) A projection of the cubic lattice in the (x,y) together with the shear mechanism illustrated by the blue arrows . These examples are for a system with size $N=8^3$.}
\label{fig:model}
\end{figure*}

This can be achieved by having each nonlinear spring $\alpha$ following a quartic potential,
\begin{equation}
E_w^\alpha = p_0^\alpha+p_1^\alpha x_\alpha+p_2^\alpha x_\alpha^2+p_3^\alpha x_\alpha^3+p_4^\alpha x_\alpha^4,
\label{eq:ew}
\end{equation}
where $x_\alpha=l_\alpha/a$, {with $l_\alpha$} the distance of two particles connected by the interaction $\alpha$ and $a$ a parameter capturing the typical local deformation of adjacent particles between $In$ and $Ac$ states. 
The above compromise can thus be satisfied {by choosing the right} prefactors $p_0^\alpha$ to $p_4^\alpha$ 
Given a nonlinear spring in the $In$ and $Ac$ states with rest length, respectively, $x_{In}$ and $x_{Ac}$, the springs are in a local minimum, if (i) 
\begin{equation}
p_1^\alpha+2 p_2^\alpha x_{Ac}+3 p_3^\alpha x_{Ac}^2+4 p_4^\alpha x_{Ac}^3=0,
\end{equation}
and (ii) same for $x_{In}$.
Only the relative value of energy is important, (iii) we can set all $p_0^\alpha$ to zero. (iv) We assume the stiffnesses of the nonlinear springs to be the same, 
\begin{equation}
p_4^\alpha=\frac{1}{2}k_w a^2,
\end{equation}
where $k_w$ is the parameter capturing the stiffness of nonlinear interactions that has a unit of spring stiffness.
(v) Finally, we define 
\begin{align*}
&p_1^\alpha (x_{In}-x_{Ac})+p_2^\alpha (x_{In}^2-x_{Ac}^2)+p_3^\alpha (x_{In}^3-x_{Ac}^3)\\ \numberthis \label{eq:dRT}
&+p_4^\alpha (x_{In}^4-x_{Ac}^4) = k_w a^2 b_w,
\end{align*}
where {$b_w$ }is a parameter capturing the energy difference between the $Ac$ and $In$ states.
We now have five equations (i-v) for the five parameters ($p_0^\alpha$ to $p_4^\alpha$) for each of the nonlinear spring $\alpha$. Hence, we can define the quartic potentials for all nonlinear springs. 

Specifically, we consider an elastic network embedded in a Face-Centered-Cubic (FCC) lattice in three dimensions (3D) as shown in Fig. \ref{fig:model}-A,B. The $In$ and $Ac$ states are connected by a shear mode, which is enabled by rotations of two rigid blocks on the left and right against each other, as shown in Fig. \ref{fig:model}-C. 
Such a model can be realized by having harmonic springs with spring stiffness $k$ in the two rigid blocks and the nonlinear springs $k_w\ll k$ at the boundary of the two.

The elastic energy of a linear (strong) spring $\alpha$ is then
\begin{equation}
E_s^{\alpha}=\frac{1}{2}k(l_{\alpha}-l_0)^2\, ,
\end{equation}
where $l_0$ is the rest length of particle separation on the lattice.
While, the energy of a nonlinear (weak) spring $E_w^{\alpha'}$ obeys Eq. (\ref{eq:ew}) with parameters solved by the particle distances in two states. 
The total energy of the system is then obtained by summing the energies over all two kinds of springs,
\begin{equation}
\label{eq:totE}
E=\sum_{\alpha}E_s^{\alpha}+\sum_{\alpha'}E_w^{\alpha'}\, .
\end{equation}

\end{document}